# Mechanism of saponite crystallization from a rapidly formed amorphous intermediate

Rogier Besselink,*[a,b] Tomasz M. Stawski,*[a,c] Helen M. Freeman,[a,d] Jörn Hövelmann,[a] Dominique J. Tobler,[e] Liane G. Benning.[a, f, g]

[a]German Research Center for Geosciences, GFZ, Telegrafenberg, 14473, Potsdam, Germany

[b]University Grenoble Alpes, University Savoie Mont Blanc, CNRS, IRD, IFSTTAR, ISTerre, 38000 Grenoble, France

[c]Bundesanstalt für Materialforschung und -prüfung (BAM), Division 1.3: Structure Analysis, Berlin, Germany

[d]School of Chemical and Process Engineering, University of Leeds, LS29JT, Leeds United Kingdom.

[e]Nano-Science Center, Department of Chemistry, University of Copenhagen, Copenhagen, Denmark

[f]Department of Earth Sciences, Free University of Berlin, 12249 Berlin, Germany

[g]School of Earth and Environment, University of Leeds, Leeds, United Kingdom

**ABSTRACT:** Clays are crucial mineral phases in Earth's weathering engine, but we do not know how they form in surface environments under (near-)ambient pressures and temperatures. Most synthesis routes, attempting to give insights into the plausible mechanisms, rely on hydrothermal conditions, yet many geological studies showed that clays may actually form at moderate temperatures (< 100 °C) in most terrestrial settings. Here, we combined high-energy X-ray diffraction, infrared spectroscopy and transmission electron microscopy to derive the mechanistic pathways of the low-temperature (25-95 °C) crystallization of a synthetic Mg-clay, saponite. Our results reveal that saponite crystallizes via a two stage process: 1) a rapid (several minutes) co-precipitation where ~20 % of the available magnesium becomes incorporated into an aluminosilicate network followed by 2) a much slower crystallization mechanism (several hours to days) where the remaining magnesium becomes gradually incorporated into the growing saponite sheet structure.

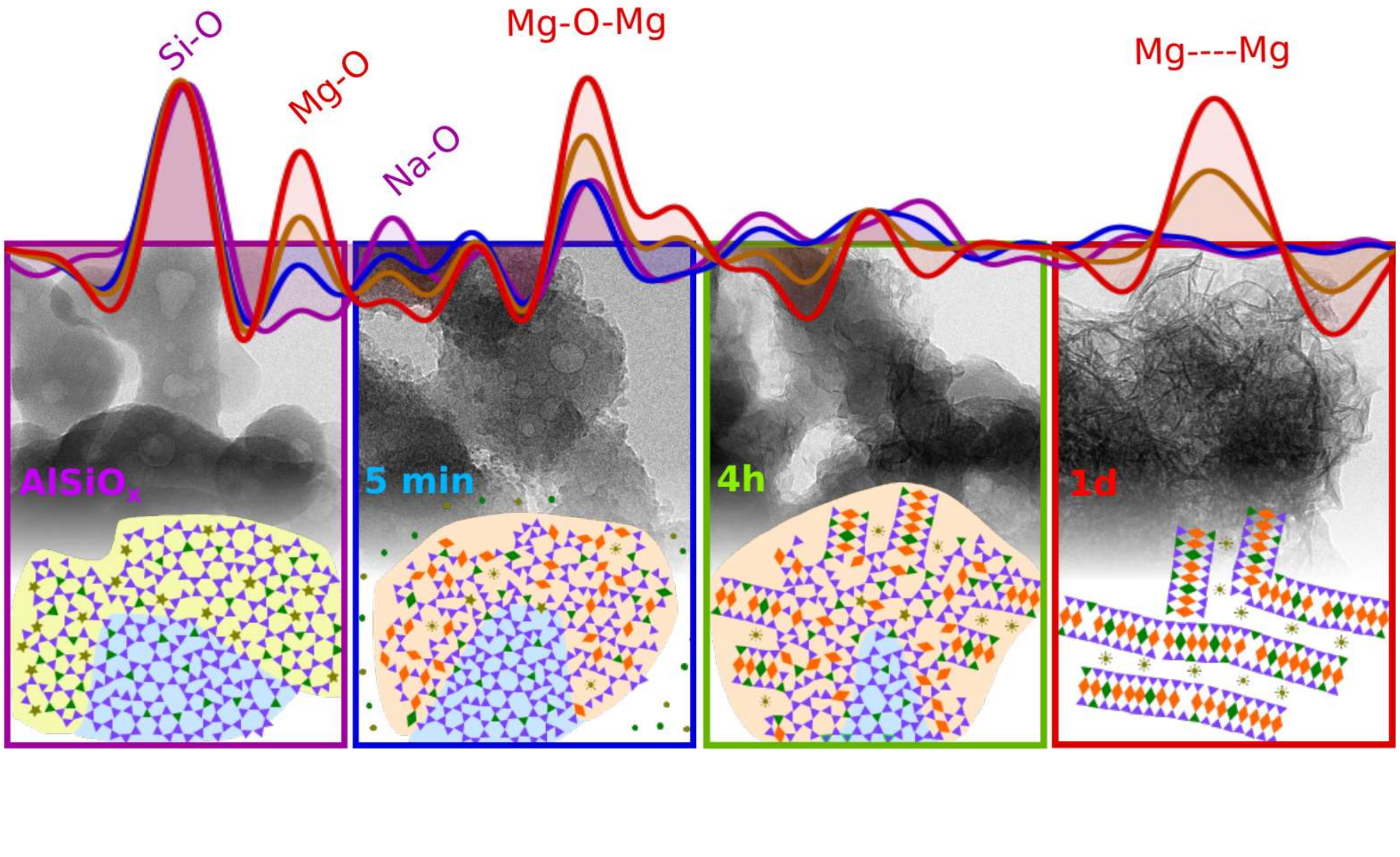

# 1. INTRODUCTION

Clay minerals are hydrous aluminum phyllosilicates rich in iron, magnesium, calcium and potassium, and are characterized by high cation exchange and high sorption capacities. Clays predominantly form during the weathering of igneous aluminosilicate minerals (e.g., feldspars, amphiboles, pyroxenes, feldspathoids and olivine), driven by fluid-rock interactions with $CO_2$-acidified waters. As such clays play an important role in global cycles of various elements e.g. iron, magnesium, potassium calcium and carbon,[1-4] and are key iron and magnesium sources for vegetation. Due to their excellent properties, clays also find wide application in industry, for example in catalysis, waste storage and confinement, paper industries, oil drilling, foundry molds and pharmaceuticals.[2, 5-6]

In this work, we focus on smectites, the most common clays in natural soils and sediments,[5] and more specifically, on saponite, the magnesium-rich trioctahedral smectite. Saponite is well-known for its low friction coefficient,[7] high surface area,[8-10] high cation exchange capacity[8-10] and catalytically active acidic surface sites.[11-13] In nature, saponite is frequently found in deep-sea hydrothermal vents, with water pressures up to 1.1 kbar. Importantly, however, together with other phyllosilicates, saponite has also been found in chondritic meteorites,[14-16] dwarf planet Ceres,[17] and on Mars,[18-19] suggesting that it can also form in water-scarce conditions and under relatively low temperature/pressure conditions (< 100 °C). However, so far, we do not fully understand how saponite forms under such geologically mild conditions.

Saponite and other smectites are typically synthesized under hydrothermal conditions,[6, 20, 21] and / or in the presence of metal alkoxide precursors (e. g. aluminum, tri-isopropoxide, and tetraethylorthosilicate); conditions that are not found in nature.[6, 20, 21] A selected number of studies successfully synthesized saponite at moderate temperatures by using inorganic salts, yet they were successful only by adding small organic chelating / templating ions or cationic metal impurities to the synthesis.[22-26] For instance, Schumann et al.[26] observed saponite formation at 60 °C in the presence of oxalic acid, which led them to suggest that organic acids are a prerequisite for clay formation in carbonaceous chondrites. Baldermann et al.[25] observed that small amount of iron impurities stabilized the saponite structure at temperatures as low as 60 °C. Noteworthy, however, that in both these studies the applied synthesis also induced brucite ($Mg(OH)_2$) formation at high pH, which constitutes parallel competitive processes to the clay formation.[25] Even though brucite may be formed in parallel under natural conditions, the presence of such impurities makes it more complex to study the saponite crystallization mechanism.

Vogels et al.[8] successfully avoided brucite formation by using a two-step synthesis approach: 1) formation of an aluminosilicate gel at high pH ≈ 13, and 2) maturation of the gel into saponite at pH 7 – 8 and 90 C. The reason for the absence of brucite in this synthesis is due to the use of urea at a temperature of 90 °C, allowing slow urea degradation to ammonia and hence controlled hydrolysis of Mg, which in turn enables gradual saponite formation while avoiding fast pH increase (i.e. brucite formation). An additional consideration is that $Al^{3+}$ ions are only 4-fold coordinated at high pH.[27] Therefore, this precursor aluminosilicate gel simulates well the 4-fold coordinated Al in aluminosilicate (e.g., volcanic glass or olivine)[2, 4] from which saponite generally forms in nature.

The synthetic route proposed by Vogels could be a plausible analog for natural saponite formation. Nevertheless, the mechanistic and structural details of the aforementioned processes are not well understood from the point of the crystal growth chemistry. For instance, it is not clear if there are any intermediate solid phases (either amorphous or nanocrystalline) involved, and what factors limit the crystallization rate. Here, we demonstrate that the saponite's growth required the formation of an intermediate phase, as is evidenced by the pair distribution function analysis on samples collected from the samples quenched at several stages of the reaction. The presence of an intermediate phase was further supported by transmission electron microscopy, which revealed the presence of amorphous spherical globules before it transformed into smectite-like sheets with well-defined interlayer distances.

# 2. MATERIALS AND METHODS

Saponite clays were synthesized following a slightly modified two-step synthesis procedure as described in Vogels et al,[8] which involved the formation of an amorphous aluminosilicate (SI Scheme S1: Step 1) followed by crystallization of saponite (SI Scheme S1: Step 2). This two-step procedure is described below and visually illustrated with a flow diagram (SI, Scheme S2). Please note that early tests indicated that the formation of saponite was most efficient when using freshly prepared alumino-silicate gel (step 1) for the synthesis of saponite (step 2). Moreover, increasing the reaction time of synthesis step 1 to 4 h (at 100 °C) led to the crystalline impurities that appeared to be less susceptible to a towards saponite formation. Therefore, each sample that is discussed here was synthesized from a freshly prepared aluminosilicate gel, which was strictly annealed for 1 h at 100 °C.

Step 1: For the synthesis of the aluminosilicate gel with [Si]:[Al]:[OH] = 6.8/1.2/6.0, the following stock solutions were prepared using reagent grade chemicals and deionized water (DIW, 18 MΩcm): $Na_2SiO_3 \cdot 5\,H_2O$ ([Si] = 1.2 M, Sigma-Aldrich >95%), $AlCl_3 \cdot 6\,H_2O$ ([Al] = 1 M, VWR, ACS) and NaOH (5 M, VWR, ACS), (see SI, section 1.1, Table S1 and Scheme S1 and S2 for more details). Inside a 5 mL polypropylene vessel with screwcap 2.10 mL of 1 M $AlCl_3$ and 2.10 mL 5 M NaOH were mixed on a vortex mixer and a precipitate was formed instantaneously according to the reaction as illustrated in SI Scheme S1: Step 1a. This precipitate dissolved within 1 min. through the formation of by the formation of $Al(OH)_4^-$, while continuing mixing on the vortex mixer. This solution was then added drop-wise with a polyethylene Pasteur pipet to 9.92 mL 1.2 M $Na_2SiO_3$ solution within a 100 mL roundbottom flask under vigorous stirring with a magnetic stirrer, at room temperature. During this addition a white precipitate was slowly formed as illustrated in SI, Scheme S1: Step 1b. The roundbottom flask was then placed in a preheated oil bath to heat the reaction mixture for 1 h at 100 °C under a reflux condenser to avoid water loss. During the reaction, both the mixture and the oil bath were continuously stirred with a combined magnetic stirrer / hotplate, with an external temperature controller that kept the oil bath temperature at 105 °C to reach a desired reaction temperature within the roundbottum flask of 100 °C, which was measured with a glass thermometer. This temperature difference was most likely caused by condensed water that drips back from the condenser back into the reactor and evaporation heat of water. Please note that the external temperature couple of the hot plate was placed inside the oil bath instead of the roundbottom flask to obtain a faster and therefore more stable temperature feedback loop.

Step 2: For the synthesis of the saponite, the following stock solutions were prepared using reagent grade chemicals and deionized water (DIW, 18 MΩcm): $MgCl_2 \cdot 6\ H_2O$ ([Mg] = 1.2 M, VWR, ACS), urea (5 M, Carl-Roth GmbH & Co, 99.5%) and hydrochloric acid (2 M, Carl-Roth, 37% p.a.). (see SI, section 1.1, Table S1 and Scheme S1 and S2). For 0.4 M histidine·HCl, 6.21 g of histidine was poured over a powder funnel into a 100 mL measuring flask. In a separate borosilicate beaker 3.94 g of HCl (37%) was poured into 40 mL DIW and this solution was poured into the flask with histidine. Subsequently, the beaker was washed three times with 10 mL of DIW and poured into the flask and the flask was filled up to its total volume. Here, histidine / hydrochloride was used as a pH buffer to avoid a fast pH decrease upon adding hydrochloric acid. As is explained in the introduction, urea was added to enable hydrolysis of magnesium by a slow release of ammonia, while avoiding a fast pH increase.

The aluminosilicate gel was cooled down until it reached a temperature below 30°C by placing the flask in a water bath (10 °C), while continuing stirring the reaction. Subsequently, it was diluted with 35.9 mL water to yield a total volume of 50 mL. In a separate bottle 8.40 mL 5 M urea, 8.75 mL 1.2 M $MgCl_2$, 8.75 mL 0.4 M histidine hydrochloride and 9 mL 2 M hydrochloric acid were all mixed and diluted with 15.1 mL DIW to a total volume of 50 mL. Then, the mixture was poured rapidly into the diluted aluminosilicate suspension under vigorous stirring. The mixture was then rapidly, but carefully neutralized to a pH in between 6.8 and 7.4, by dropwise adding 2 M of hydrochloric acid. (this corresponded roughly to 1.5 mL of 2M hydrochloric acid, but the amount seem to vary depending on the quality of the aluminosilicate gel). Then, this reaction mixture was placed inside a preheated oil bath (110 °C) to reach a desired reaction temperature of 95 °C. For this purpose we have used the same experimental setup as for step 1, with the oil bath placed on top of a combined magnetic stirrer / hot plate and an external temperature couple placed inside the oil bath. Due to the fact that the aluminosilicate suspension is not stable at pH 7 and 25 °C, it was important to execute the mixing and pH neutralization within ~10 min, and it need to placed inside a hot oil bath immediately afterwards (i.e. the gel had to be fresh since $Al^{3+}$ will slowly dissolve and becomes 6-fold coordinated).

Once the reaction mixture reached the desired temperature (95 °C ~ 15 min), the time was set to 0 and the crystallization reaction held at this temperature for various time durations (5 min to 90 d). Afterward, the suspensions were cooled to room temperature and then centrifuged (10 min at rcf ≈ 9.5 kG) to isolate the solids. The supernatant was discarded and the solids re-suspended in DIW. This washing procedure was repeated 3 times to remove any residual salts. The majority of the solid paste was placed on a petri dish into a desiccator to dry for 5 d (10-30 mbar, 25 °C). For glycolization, selected dried solids and glycol were placed inside a desiccator kept at 60 °C and ~ 10 mbar. By using the same approach, samples were also synthesized at 60 °C and 25 °C. In a separate set of experiments the Mg concentration was increased to yield a [Mg]:[Si] = 12:6.8, while keeping the total volume constant.

The synthesized samples were characterized with, transmission electron microscopy (TEM; equipped with energy dispersive X-ray spectroscopy (EDS)), Fourier transform infra-red spectroscopy (FTIR), X-ray diffraction (XRD), high-energy X-ray diffraction (HEXD) and the cation exchange capacity was measured by methylene blue adsorption. All details are described in the SI: Supporting Materials and Methods.

## 3. RESULTS AND DISCUSSION

Saponite was synthesized in two separate reaction steps, i.e. (Step 1) the formation of an amorphous aluminosilicate network, (Step 2) the crystallization of the amorphous network towards saponite in the presence of magnesium and urea. Our preliminary observations of the reaction in Step 2 suggested that the amorphous aluminosilicate gel reacted very fast at its surface with magnesium, but it did not immediately crystallize to saponite. This reaction step was therefore subdivided into two substages.

### 3.1. Synthesis step 1: The formation of an amorphous aluminosilicate gel

The gel that formed immediately after mixing the silicate and aluminate solutions, was composed of spheroidal aggregates/particles or globules ranging in diameter from 50 to 200 nm as revealed by TEM imaging (Fig. 1a). These aggregates did not contain any sharp edges or lattice fringes, which may suggest that the material is possibly amorphous. This was also confirmed by XRD which showed a broad maximum at d = 3.09 Å (Cu-K$\alpha$ $2\theta$ = 28.9°, α in Fig. 2, pattern I) stemming from the Si-Si and Si-Al distances common in amorphous aluminosilicates with dihedral Si-O-Si angles of 150° and the Si-O and Al-O distances close to 1.60 Å.[28-29] The auxiliary FTIR spectrum contains both in-phase and out-of-phase antisymmetric Si-O vibrational modes (SI: Fig. S1, Table S1), which is consistent with an amorphous aluminosilicate as discussed in more detail in the SI. Moreover, in TEM imaging these structures typically exhibited heterogeneities, with the cores being more electron transparent (brighter) compared to the rims. EDS analysis indicated that the brighter regions were depleted in Na compared to the darker rims (SI: Table S2). Furthermore, the aggregates overall were depleted in Al with respect to the initial Al: Si ratio (1.2:6.8), indicating that the available Al was not stoichiometrically incorporated into the precursor aluminosilicate gel.

### 3.2. Synthesis step 2, sub-stage 1: Reaction of magnesium at the surface of the aluminosilicate.

Immediately after pH neutralization and $MgCl_2$ addition but prior to heating, small nanosized particles were formed at the surface of the aluminosilicate particles as revealed by TEM imaging (< 20 nm, Fig. 1b). In addition, a new broad reflection was observed in the XRD pattern at d = 3.77 Å, (Cu-K$\alpha$ $2\theta$ = 23.6°, β in Fig. 2, pattern III). This reflection corresponds to distances larger than Si-O-Si/Al bonds and is best explained by the formation of Mg-O bonds, ~2.05 Å.[33] This is also supported by the fact that this reflection was absent in a sample of the aluminosilicate gel which was neutralized but not mixed with the $MgCl_2$ solution (Fig. 2, pattern II).

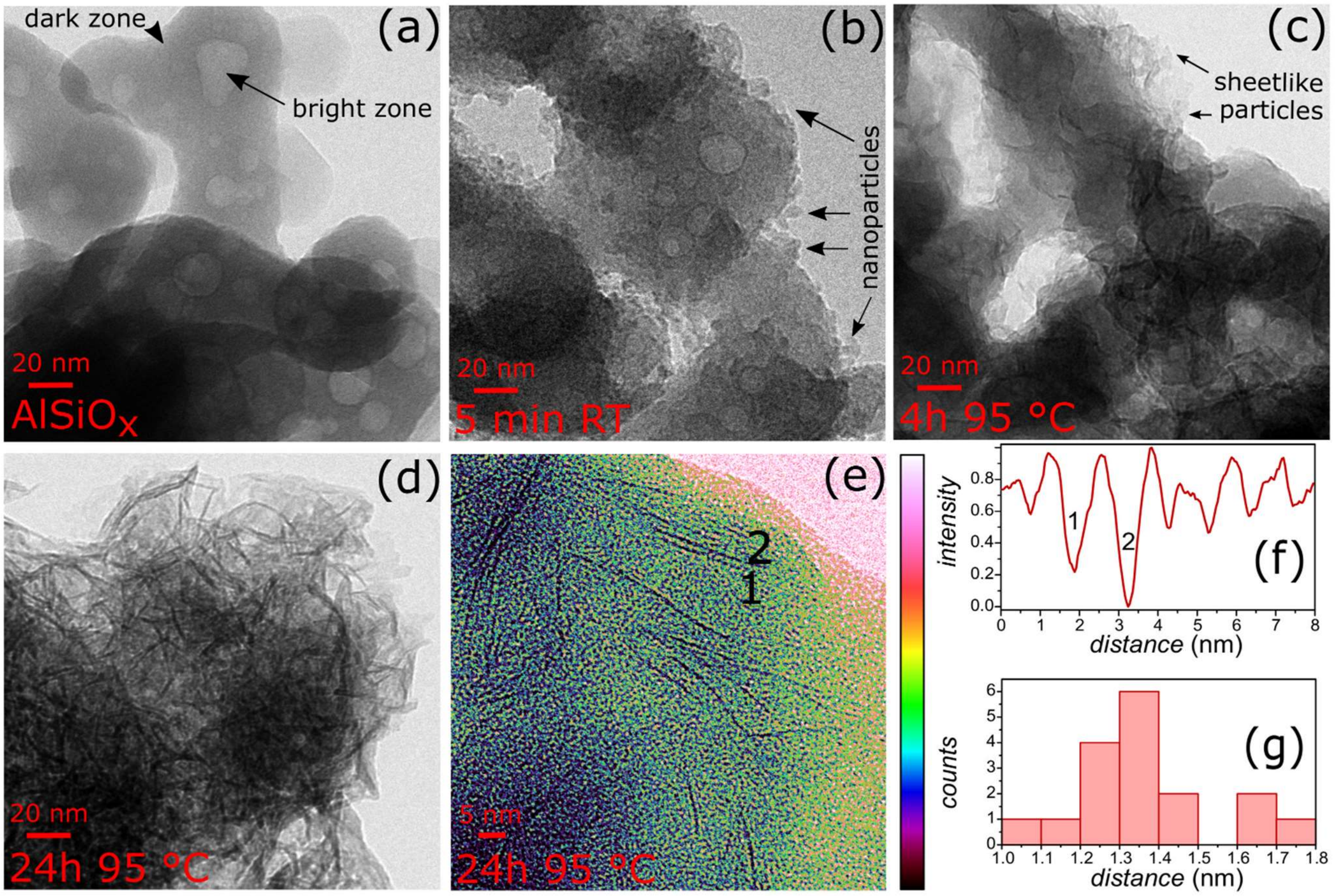

**Fig. 1.** (a)-(d) Bright field TEM images of solids removed at different sub-stages during saponite crystallization: (a) precursor aluminosilicate gel; (b) 5 min after mixing $MgCl_2$ and aluminosilicate gel at 25 °C; (c) after 4 h crystallization at 95 °C; (d) after 24 h crystallization at 95 °C. (e) TEM image (with false-color intensity scaling) of 24 h/95 °C saponite dispersed in cured LR-white, showing saponite stacks; (f) distance profile across saponite sheets indicated by numbers 1, 2 in (e); (g) histogram of distances between sheets from 17 different particle stacks, which indicates an average interlayer distance of 1.3 nm.

### 3.3. Synthesis step 2, sub-stage 2: Crystallization of saponite.

After 4 h of reaction at 95 °C, the nanoparticle – gel mixture still contained mostly spherical aluminosilicate aggregates but in addition sheet-like particles also appeared (Fig. 1c). After another 20 h (Fig. 1d), the original morphology fully disappeared and only the sheet-like particle morphology persisted. Thus, during this period, amorphous aluminosilicate gradually transformed into a layered structure as evidenced by the 24 h/95 °C sample. We embedded this product in an acrylic resin, which allowed us to visualize and measure distances between individual sheets[26, 30] (Figs. 1 e, f). The average sheet-to-sheet distance (as determined from 17 different stacks) was 1.37 ± 0.37 nm (Fig. 1g), which is consistent with typical interlayer distances for smectite clays.[26, 31]

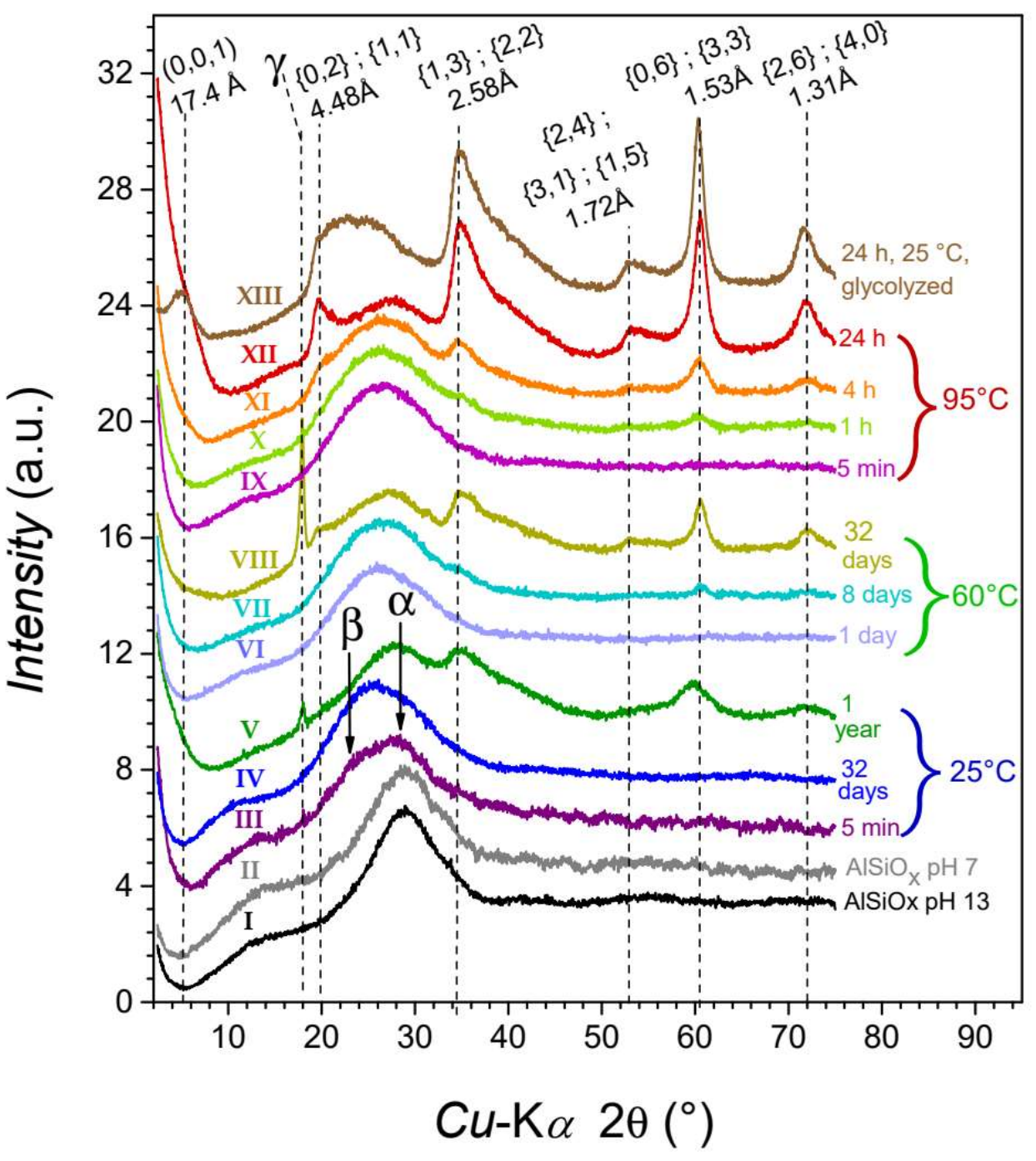

**Figure 2.** XRD patterns of synthesis products from the various substages of the crystallization step at 5, 60, and 95 °C; the reflections with α and β refer to correlation peaks originating from metal-metal distances, i.e., α relates to tetrahedral metal cation distances ($^{IV}$Si, $^{IV}$Al) and β relates to octahedral metal cation distances ($^{VI}$Mg, $^{VI}$Al); the peak marked with γ corresponds to a gibbsite impurity in the long-term synthesis products at 25 °C and 60 °C.

The transformation of the material towards a smectite-type clay was also evidenced by the appearance and progressive growth of smectite-like reflections in XRD patterns (Fig. 2, patterns IX-XIII).[24, 32-33] The asymmetric nature of the peaks with *hk*-indices {1,3}; {2,2} and {2,4}; {3,1}; {1,5} are indicative of substantial turbostratic disorder, which involves both translational and rotational disorder in the stacking of the sheets on top of each other.[34]

Meanwhile, the FTIR spectrum (SI: Fig. S1, Table S2) reveals the presence of the (Mg)OH liberation mode that is associated with a trioctahedral clay structure and narrowing of the antisymetric SiO vibration that is associated with a reduced disorder of the silica network (as discussed in more detail in the SI). The (0,0,1) reflection was only observed for the glycolized sample (Fig. 2, XIII), at $d$ = 17.4 Å (Cu K$\alpha$ 2$\theta$ = 5.3°), and matched the interlayer spacing of smectite-like clays with an interlayer charge $z$ < 1.2 per $[Mg;Al]_6$ $[Si; Al]_8O_{20}(OH)_4$ unit.[33-34] The position of the {0,6}; {3,3} reflection, corresponding to a spacing of $d$ = 1.53 Å, is consistent with the interlayer spacing of a trioctahedral clay structure.[33-34]

At lower synthesis temperatures (25 and 60 °C) similar smectite type features were formed, however, the reaction kinetics were much slower. This is shown by comparison of the XRD patterns in Fig. 2: the 4-hour pattern at 95 °C matched the 32-day pattern at 60 °C and the 1-year pattern at 25 °C in terms of peak location and intensity (all measured by using the same measurement procedure, and for same amount of solid). The only visible difference was a diffraction peak at 2θ = 17.9°, stemming from gibbsite (4.82 Å, (0,0,2)), that emerged in samples after a long reaction time ($\gamma$ in Fig. 2, V and VIII). Note that no other secondary phases (e.g. talc and chlorite interstratifications) were observed. Thus, we can confirm that saponite can form at ambient temperatures (< 95 °C).

### 3.4. Detailed evolution of the local atomic structure upon saponite crystallization.

More detailed information regarding the local environment of atoms was obtained from the high energy X-ray diffraction (HEXD) measurements (SI: Fig. S2) and their PDF analysis (Fig. 3). The PDF of the final product after 19 d at 95 °C (Fig. 3a, black pattern labeled with (I)) can be compared with simulated saponite profiles (Fig. 3a II-IV) from chemically adjusted crystallographic data of its structural analogue hectorite (Fig. 3b).[35] In these PDF profiles, the distance at 1.63 Å is associated with T-O distances, where T represents Al and Si cations tetrahedrally coordinated by oxygen.[35] This characteristic distance is typical for silicates and was found in all our samples.

The decrease in correlation intensity with increasing distance, $r$ was more evident for the experimental data compared to the simulated PDF of bulk saponite crystal. To a large extent, this can be explained by the fact that the synthesized saponite consisted of nanocrystals and this effect is largely reproduced in the simulated profiles by introducing a limited particle size diameter of ~25 Å that suppressed higher distance correlations (Fig. 3a III). On the other hand, many distance correlations of the experimental PDF were also present in the simulated curves (Fig. 3a II-IV) except for the correlations at 2.38 and 3.66 Å. The distance at 2.38 Å matched with Na-O distances pairs in crystalline tectosilicates (e.g., hydrosodalite, pitiglianoite and cancrinite[35]) and its intensity substantially decreased during the transformation into more crystalline silicates (Fig. 3c). Based on TEM-EDS analyses (SI: Table S3) roughly half of the fraction of sodium ions leached out within the first 5 min of reaction, which likely explains the first decrease in Na-O pair intensity.

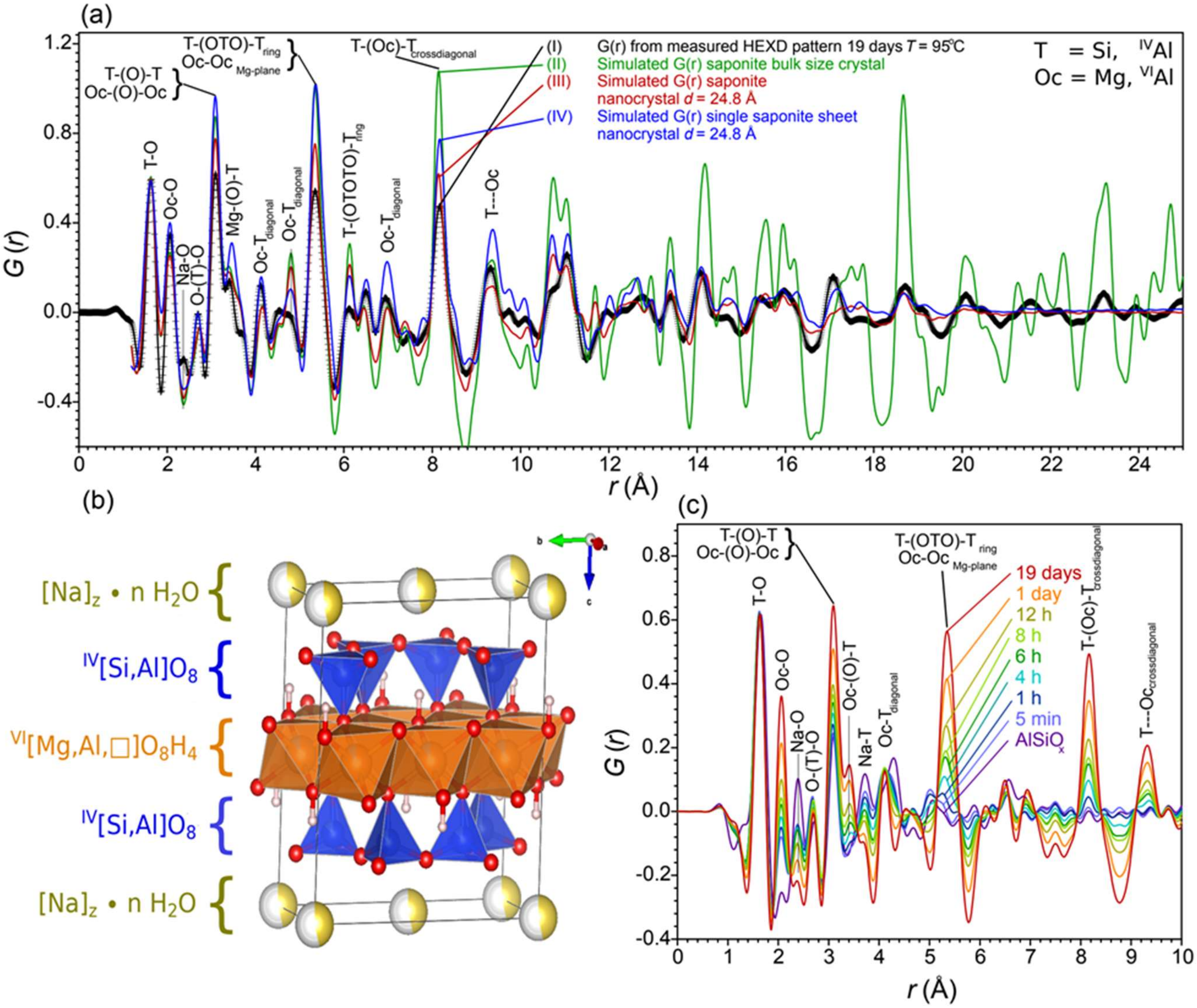

**Figure 3.** (a) PDFs of I) samples reacted for 19 d at 95 °C (see more info also in SI, Fig. S2); II) simulated G(r) of a saponite structure,[33] with a modified stoichiometry (from SI: Table S3), and with cell parameters and anisotropic displacement parameters optimized to the experimental data, while keeping relative atom positions constant; III) simulated G(r), nanocrystal size damping ~25 Å of curve (II); IV) simulated PDF pattern of a saponite sheet by placing the saponite sheet from II in a unit cell with an extended c-axis and consequently eliminating cross-correlations from neighboring sheets and sodium ions. (b) Simulated generalized type-II trioctahedral saponite structure[41]; (c) PDFs from measured total scattering curves of samples reacted at 95 °C for between 5 min to 19 d, including the PDF pattern of the precursor aluminosilicate gel. All simulated and measured PDF patterns were normalized against the T-O peak intensity of the 19 d/95 °C saponite

At longer reaction times, the tectosilicate-type Na-O pair intensity (at 2.38 Å) further decreased, whereas TEM-EDS indicated that the Na fraction remained constant after 24 h (SI: Tables S3 and S4). Consequently, at t > 24 h the Na-O pair intensity likely decreases due to the rearrangement of sodium ions into different lattice positions. Moreover, Na-O pair distances are probably larger for phyllosilicates as it was observed for instance for nontronite with Na-O pairs at distances in a range between 2.8 and 3.4 Å.[35] Thus, the overall decrease in Na-O intensity can be explained by two phenomena: at short reaction times, sodium predominantly leaches out of the aluminosilicate gel, whereas after longer reaction periods sodium ions rearrange into interlayer position.

The peak at 3.66 Å (Fig. 3c.) most likely relates to Na-T pairs despite being larger than the expected distance in tectosilicates (d(Na-T) = 3.40 Å).[35] However, the Na-O-T bond angles (∠(Na-O-T) ≈ 131°) in disordered framework silicates are larger compared to well crystalline tectosilicates (∠(Na-O-T) ≈ 115°), which in turn could explain the larger Na-T distance. The pair distance correlation at 2.68 Å corresponds to O-(T)-O distances for both tecto- as well as phyllosilicates, and these intensities remain roughly constant throughout the crystallization reaction. Finally, pair distance correlations at 2.06 and 3.38 Å are associated with Oc-O bonds and Oc-(O)-T distances of the modified saponite structure, where Oc represented octahedral metal cations, Mg and $^{VI}Al$. The intensity of these distance correlations increases with increasing reaction time, which is likely a consequence of condensation of magnesium and octahedral aluminum into the aluminosilicate framework. Based on the simulated structures, next-nearest neighbors were assigned to correlations at given distances (Fig. 3a), which could be subdivided into in-plane correlations at 3.04, 5.34 and 6.11 Å, and cross-diagonal plane correlations at 3.47, 4.09, 4.78, 6.10, 6.99, 8.18 and 9.36 Å.

The slight disagreement between the experimental and the simulated PDF of saponite nanocrystals (Fig. 3a III) indicates structural imperfections in the synthesized saponite (Fig. 3b). This might be due to turbostratic disorder (i.e., faults in the stacking of the sheets). To verify this we calculated the PDF of an isolated saponite sheet with a c-axis extended to 100 Å, while keeping absolute atom positions fixed (Fig. 3a IV). The juxtaposition with our data showed that most of the observed distance correlations in the measured data match the ones from the simulated single sheet, indicating that our data do not actually exhibit correlations across sheets, thus it is unaffected by the stacking of the sheets. Consequently, the reduced correlations in our data (Fig. 3a I) when compared to the simulated sheet pattern (Fig. 3a IV) cannot fully be explained by turbostratic defects or reduced crystal domain sizes. Thus, the intensity differences between data and model may also be due to defects within the sheet structure, such as a ditrigonal distortion, cation substitution and/or cation vacancies as are also seen in other clay systems.[33, 36-38]

The presence of defects due to cationic substitution was evaluated through the compositional analysis from TEM-EDS and by solving charge and mass balances equations (see SI: Supporting Data Analyses). Consequently, the fractions of 4-fold-coordinated $^{IV}Al^{3+}$ and 6-fold-coordinated $^{VI}Al^{3+}$ could be estimated. Both the 24 h and the 19d samples had a reduced rMg ratio and an enlarged rSi ratio compared to the idealized composition: $Na_{1.2}[Si_{6.8}{}^{IV}Al_{1.2}][Mg_6]O_{20}(OH)_4$ (SI: Table S4). Moreover, the Na content and the cation exchange capacity (CEC) were lower compared to the ideal composition, which suggested a reduced interlayer charge. This can be expected when aluminum partially occupies the octahedral position ($^{VI}Al$) and forms a type-II saponite as illustrated in SI: Scheme S3b. The 19 d sample has an increased $^{IV}Al$ concentration in comparison to the 24h sample, which suggests slow incorporation of additional $Al^{3+}$ ions on tetrahedral sites. As $Al^{3+}$ ions become incorporated in the tetrahedral network, the clay structure becomes more negatively charged as evidenced by an increased CEC (0.15 to 0.42 mmol/g) and $Na^+$ fraction (0.16 to 0.33 mmol/g) when going from 24 h to 19 d (SI: Table S4).

### 3.5. Reaction Kinetics and proposed mechanism of saponite formation.

The time-dependent change in XRD peak intensities was used to estimate the rate of saponite crystallization for each tested synthesis temperature (25, 60 and 95 °C, Fig. 4a). The reaction progress was best described by the first-order reaction rates (Fig. 4a and SI: Table S4). Since the reaction kinetics were similar for two different magnesium concentrations (Fig. 4a. 95 °C, $Mg_6$ and $Mg_{12}$), we can clearly state that the condensation of magnesium into the aluminosilicate network is not the rate-limiting process (see also SI: Supporting Data Analyses).

Time-dependent change in the PDF normalized peaks intensities were used to evaluate structural changes in the crystallization process. Changes in the intensities of the Na-O (2.38 Å), Oc-O (2.06 Å) and T-(Oc)-T (8.18 Å) pair correlations (Fig. 4b) revealed that the amount of Na-O and Mg-O distances changed rapidly within the first 5 min. In contrast, the T-(Oc)-T pair at 8.18 Å only reveals a significant change after ~ 1 h of reaction similar to the other longer-range correlations (Fig. 3b). These trends indicate that the ordering and crystallization reactions proceed via a two-stage-process. The first stage involves a fast re-structuring of short-range distances, where $Na^+$ ions rearrange and $[Mg^{2+}\cdot n\ H_2O]$ condense into a disordered aluminosilicate network, which is completed within the first 5 min at 95 °C. During this stage, 30% of the Na-O bonds remained at their original tectosilicate-like distances from oxygen, while 70% became rearranged within the solid and/or dissolved (Fig. 4b), TEM-EDS analysis further showed that only ~ 50% of the Na ions actually remained present in the solid material (SI: Table S3), i.e., ~50% of the Na ions dissolved. As such, only 20% rearranged to different lattice positions.

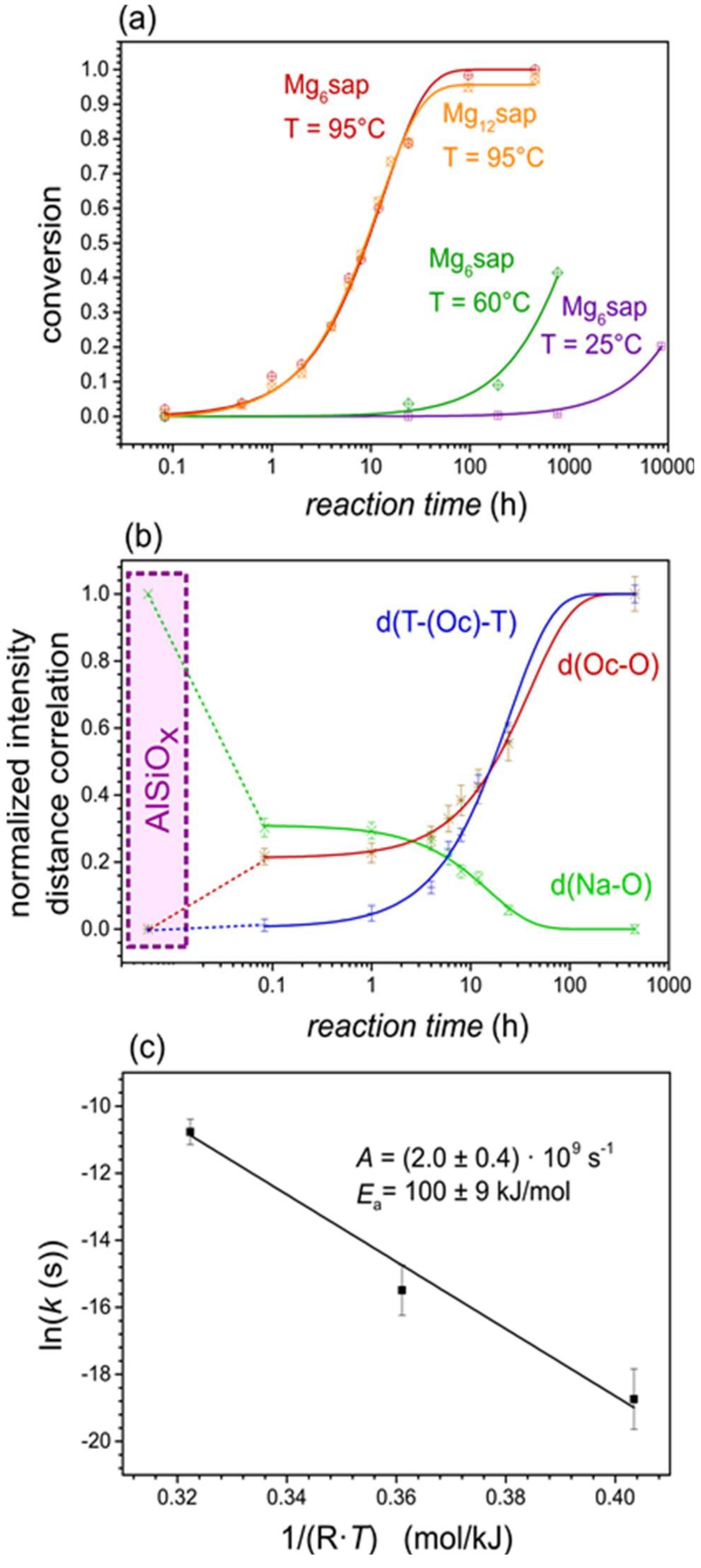

**Figure 4.** (a) Saponite formation rates based on overall peak intensities of powder XRD at 25, 60 and for 95 °C at two magnesium ratios, 6 and 12 per $O_{20}(OH)_4$, referred to as $Mg_6$sap (stoichiometric) and $Mg_{12}$sap (double the stoichiometric amount); (b) α and reaction progress based on distance correlations at 2.04, 2.38 and 8.18 Å corresponding to Mg-O, Na-O and T-(Mg)-T distances respectively; (c) Arrhenius plot based on reaction rates from the evaluated XRD intensities in (a)

The second stage is much slower and affects both short- and long-range atomic distances, which eventually leads to the formation of a layered sheet-like structure, with the same symmetry constraints as hectorite. During the whole process, ~20% of the Oc-O pairs was formed during the first stage and the remaining ~80% during the second stage (Fig. 4b). Meanwhile, only ~15% of the $Mg^{2+}$ fraction within the 19 d sample was condensed to the aluminosilicate network within the first 5 min. The ~5%-difference between the Oc-O pairs and the condensed $Mg^{2+}$ fraction is likely explained by octahedral $^{VI}Al$ ions that contribute to the measured pair distance at 2.06 Å. Furthermore, the Mg-O condensation and rearrangement of $Na^+$ ions were likely limited by the slow rate of the sheet formation as indicated by the delayed increase of the T-(Oc)-T correlation. After 24 h at 95 °C, our TEM-EDS data revealed an increased $^{IV}Al^{3+}$ and $Na^+$ ion concentration in the synthesized saponite (SI: Table S4) Thus, as the fraction of the incorporated $^{IV}Al^{3+}$ increases, the interlayer charge increases, allowing more $Na^+$ ions to diffuse into the interlayer of saponite, which stabilizes its structure.

From the temperature dependent reaction rate constants that were derived from the conversion diagrams (Fig. 4a, SI: Table S5) we also estimated a crystallization activation energy of ~100 kJ/mol (Fig. 4c). Similar activation energies were found for the transformation and hydrolysis reactions in varying types of layered silicates, including: beidelite to illite/smectite transformation,[39] Na-montmorillonite hydrolysis,[40] kaolinite to illite,[41] and the kaolinite to mullite transformations.[42] The activation energy in the range from 80 to 120 kJ/mol was associated with the disruption of Si-O bonds.[43] Our results suggest that the fraction of $^{IV}Al^{3+}$ increases during the progressive saponite crystallization and concidering that Al-substitution involves disruption of silicate bonds, this Si-O-bond disruption may be one of the key rate-limiting processes in saponite formation. On the other hand, diagenetic smectite to illite transformation reactions are often described as requiring lower activation energies (in the range of 37-70 kJ/mol)[44] and as such those transformations are limited by ion exchange processes and not a disruption of Si-O bonds.

### 3.6 Proposed mechanism of saponite formation

Based on all the data discussed above, we can summarize the process of saponite crystallization (Fig. 5). Here, a precursor aluminosilicate gel is formed in Step 1 and is composed of mixed silica/alumina tetrahedra (Figure 5a: ▼, ▼) and interconnected sodium ions (★), which are predominantly located within the outer (rim) regions of the gel. The gel is mixed with magnesium ion, which starts Step 2 of a multi-stage reaction:

Step 2, sub-stage 1: <5 minutes ~15 % of the magnesium ions condense (Fig. 5b, ♦) within the aluminosilicate framework. From the available Al ions, 50% dissolve (•), 27% rearrange into an octahedral coordination (♦) and 23% remain in the tetrahedral coordination (▼); ~ 50 % of the strongly interconnected sodium ions (Fig. 5a, b: ★) leach into solution (•) and ~20% rearrange to different lattice positions (☀).

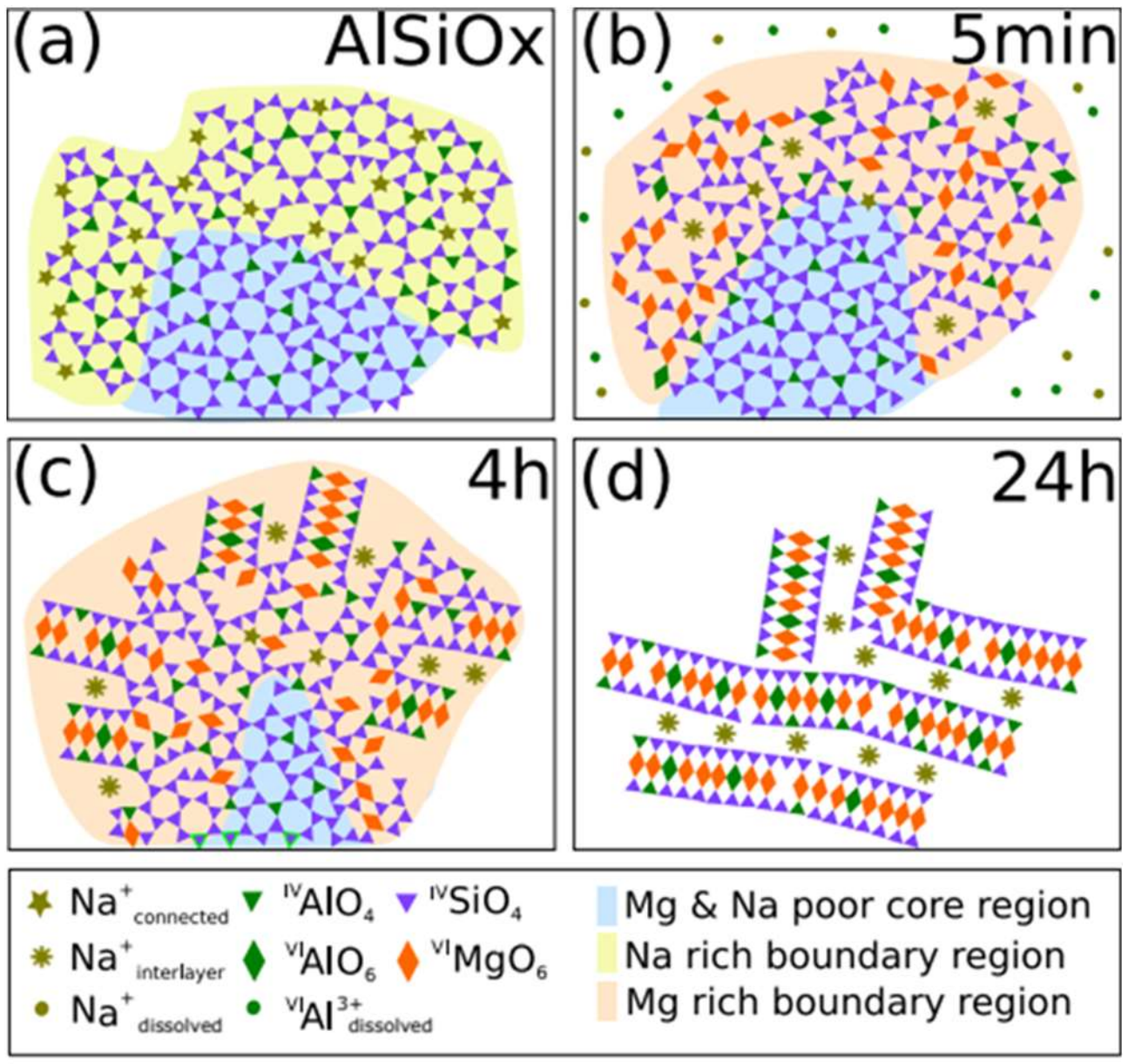

**Figure 5.** Schematics of the saponite crystallization: (a) AlSiOx precursor gel; (b) amorphous Mg-rich aluminosilicate gel, c) after 4 h at 95 °C with initial layered materials growing at the boundary; (d) after 24 h with the majority of material transformed to saponite.

Step 2, sub-stage 2: at a slower rate, the atoms within the amorphous aluminosilicate slowly rearrange towards a brucite-like layered trioctahedral sheet structure and the remaining magnesium ions become incorporated within the structural framework. Rather counter-intuitively, this incorporation of magnesium is not the rate-limiting step. The crystallization is in fact limited by either 1) disruption of the Si-O bonds which enables the incorporation of additional $^{IV}Al^{3+}$ ions (▾) and/or by 2) diffusion-limited separation of the octahedrally coordinated Mg and Al cations (♦,♦) from the tetrahedrally coordinated Si and Al cations (▾, ▾, Fig. 5b-d). As soon as the structure is reorganized into such sandwich-like T-O-T fragments, these fragments may continue to grow in the lateral dimension through nearly oriented attachment of such poorly crystalline fragments as suggested by a recent agon absobtion/STEM characterization experiments on saponite.[42] Consequently, even after 19 d at 95 °C, the saponite sheets still contained many defects (as evidenced in the calculated next-neighbor correlations from PDF data). Based on this proposed mechanism and the derived activation energy and reaction rates, we estimate that at 25 °C saponite would require at least 14 years to convert 95% of a precursor amorphous aluminosilicate to a trioctahedral saponite clay.

In overall, these results suggests that saponite forms through multi-nucleation pathway via amorphous or highly disordered intermediate phases as it was reported for several other mineral systems,[43-46] which may have two causes: Firstly, the formation of stable saponite relies on the fraction of aluminum that stays in the fourfold coordination. The aluminum ions that dissolve near neutral pH, become sixfold coordinated,[27, 47] which likely leads to a higher fraction of octahedral vacancies as illustrated by cation substitution SI Scheme S3. Moreover, Bisio[48] had shown that if saponite was synthesized from a more hydrated aluminosilicate gel $H_2O/Si = 50$ as compared to $H_2O/Si = 20$, a less crystalline saponite was formed, with an increased fraction of 6-fold coordinated aluminum based on $^{27}Al$-NMR. Thus, saponite may need to be formed through solid state transitions to keep a sufficient amount of aluminum in the fourfold coordination. Secondly, the interfacial energy of saponite may be too high for saponite to be formed through a single-step nucleation process as it was suggested for other mineral systems.[43-46]

## 4. CONCLUSION

Our results clearly document how saponite forms from a precursor aluminosilicate gel via an amorphous intermediate Mg-rich phase where 15% of the magnesium fraction reacts within 5 min at 95 °C. This is followed by a slower crystallization to saponite, which takes more than 8 h at 95 °C. The magnesium concentration essentially does not affect the rate of crystallization but the structural arrangement to sheets seems to be the rate-limiting step in the crystallization reaction as evidenced by the delayed increase in sheet-related distance correlations compared to the Mg-O and $^{VI}Al$-O distances. Hence, the process proceeds via a very fast formation of an Mg-rich amorphous intermediate that subsequently slowly reorganizes into a sheet-like structure of saponite.

## ASSOCIATED CONTENT

### Supporting Information

1. Supporting Materials and Methods:
   1.1. Synthesis of saponite, (Table S1, Scheme S1 and S2)
   1.2. Transmission Electron Microscopy (TEM)
   1.3. X-ray diffraction (XRD, HEXD)
   1.4. Fourier Transform Infra-Red (FTIR) spectroscopy.
   1.5. Cationic Exchange Capacity (CEC).

2. Supporting results and data analysis:
   2.1. Assignment of FTIR vibration modes during various stages of the saponite formation (Figure S1 and Table S2)
   2.2. Compositional analysis, mass balances and cationic exchange capacity (Scheme S3, Tables: S3 and S4).
   2.3. High Energy X-ray Diffraction Patterns, (Figure S2)
   2.4. Kinetics of saponite formation (Table S5)

## AUTHOR INFORMATION

### Corresponding Authors

* Email: besserogier@gmail.com, tomasz.stawski@gmail.com, benning@gfz-postdam.de

**ORCHID:**
R. Besselink: 0000-0002-2027-9403
D. J. Tobler: 0000-0001-8532-1855
T.M. Stawski: 0000-0002-0881-5808
H.M. Freeman 0000-0001-8242-9561
L.G. Benning 0000-0001-9972-5578

**Notes**
The authors declare no competing financial interest.

## ACKNOWLEDGMENT

This research was funded by a Helmholtz Recruiting Initiative grant which supported R.B., T.M.S., H.M.F., J. H. and L.G.B. This research was also supported by a Marie Curie grant from the European Commission in the framework of the NanoSiAl Individual Fellowship, Project No. 703015 to T.M.S.
We also want to thank Karina Chapman, and Kevin A. Beyer for assistance with HEXD analyses at the APS beamline 11 ID-B. Use of the Advanced Photon Source was supported by the U. S. Department of Energy, Office of Science, Office of Basic Energy Sciences, under Contract No. DE-AC02-06CH11357. We would like to thank our GFZ colleagues for technical assistance with the XRD (Anja M. Schleicher, TEM sample preparation (Richard Wirth and Anja Schreiber) and FTIR (Ilias Efthimiopoulos and Monika Koch-Müller).

*Supporting Information for:*

# Mechanism of saponite crystallization from a rapidly formed amorphous intermediate

Rogier Besselink,*[a,b] Tomasz M. Stawski,[a,c] Helen M. Freeman,[a,d] Jörn Hövelmann,[a]
Dominique J. Tobler,[e] Liane G. Benning.[a, f, g]

[a]German Research Center for Geosciences, GFZ, Telegrafenberg, 14473, Potsdam, Germany
[b]University Grenoble Alpes, Univiversity Savoie Mont Blanc, CNRS, IRD, IFSTTAR, ISTerre, 38000 Grenoble, France
[c]Bundesanstalt für Materialforschung und -prüfung (BAM), Division 1.3: Structure Analysis, Berlin, Germany
[d]School of Chemical and Process Engineering, University of Leeds, LS29JT, Leeds United Kingdom.
[e]Nano-Science Center, Department of Chemistry, University of Copenhagen, Copenhagen, Denmark
[f]Department of Earth Sciences, Free University of Berlin, 12249 Berlin, Germany
[g]School of Earth and Environment, University of Leeds, Leeds, United Kingdom

## Contents:

# 1. Supporting Materials and Methods

## *1.1. Synthesis of saponite:*

**Table S1.** Recipes stock solutions

| Synthesis step | compound | supplier | purity | Concentration (mol/L) | Mass compound (g) | Measuring flask volume (mL) | Measuring flask type |
|---|---|---|---|---|---|---|---|
| 1a | NaOH | VWR | ACS | 5.0 | 20.00 | 100 | PFA* |
| | $AlCl_3 \cdot 6\ H_2O$ | VWR | ACS | 1.0 | 24.14 | 100 | borosilicate |
| 1b | $Na_2SiO_3 \cdot 5\ H_2O$ | Sigma-Aldrich | > 95% | 1.2 | 63.64 | 250 | PFA* |
| 2 | $MgCl_2 \cdot 6\ H_2O$ | VWR | ACS | 1.2 | 60.99 | 250 | borosilicate |
| | urea | Carl-Roth | 99.5% | 5.0 | 75.22 | 250 | borosilicate |
| | Histidine<br>HCl (37%) | VWR<br>Carl-Roth | 99%<br>p.a. | 0.4<br>0.4 | 6.21<br>3.94 | 100 | borosilicate |
| | HCl (37%) | Carl-Roth | p.a. | 2.0 | 49.3 | 250 | borosilicate |
| *PFA = perflouroalkoxy alkane | | | | | | | |

**Preparation of stock solutions:**

Before starting the synthesis stock solutions were prepared according to Table S1. Because of the high concentration that were required for most compounds all compounds except for 37% hydrochloric acid were directly mixed inside the measurement flasks. Moreover, to avoid potential corrosion the measuring flasks, the stock solutions of the alkaline compounds: NaOH and $Na_2SiO_3 \cdot 5\ H_2O$ were prepared in PFA (perflouroalkoxy alkane) measuring flasks. The stock solutions of NaOH, $AlCl_3 \cdot 6\ H_2O$, $Na_2SiO_3 \cdot 5\ H_2O$, $MgCl_2 \cdot 6\ H_2O$ and urea were prepared by pouring the specified amount of weighted solid over a polyethylene powder funnel into a cleaned and dried measuring flaks. Then, the powder funnels was washed three times with 5 - 10 mL of deionized water (DIW) to remove any solid form the powder funnel and the neck of the measuring flask. Then the flask was filled up to ~75 % of its total volume, a stopper was placed on top and the flask was placed for ~ 2 min in an ultrasonic bath. Afterwards, the flaks was swung vigorously in a circular horizontal motion by hand for ~ 30 s obtain a better intermixing of the between the top part and bottom part of the mixture. (Please note that this swinging procedure is substantially more effective if the flask is not completely filled up until its total volume and the flask was swung horizontally to avoid getting particles in the neck of the flask). Then, the flask was placed back into the ultrasonic bath again and this cycle of ultra-sonication and swinging the solutions was repeated several times until all solids were visually dissolved. Only when all solid was completely dissolved, the flaks was filled up to its total volume.

For 2M HCl, 49.3 g of concentrated acid was poured into 150 mL DIW inside a borosilicate beaker. This solution was then transferred into a 250 mL measuring flask. The beaker was then washed three times with 10 mL of water and transferred into the measurement flask and filled up until its final volume.

For 0.4 M histidine·HCl, 6.21 g of histidine was poured into a 100 mL measuring flask. In a separate borosilicate beaker 3.94 g of HCl (37%) was poured into 40 mL DIW and this solution was poured into the flask with histidine. Subsequently, the beaker was washed three times with 10 mL of DIW. Then, the histidine was fully dissolved following the same ultra-sonication / swinging routine as for the other solutions.

Then all the solutions were left standing for 30 min. to cool down to room temperature, since the temperature for most solutions had increased the due to ultra-sonication and dissolution heat of these compounds. Then the flasks were filled up a second time until its total volume and homogenized by flipping the flasks upside down for 20 times. The solutions were then poured into dry polyethylene bottles and used for several syntheses.

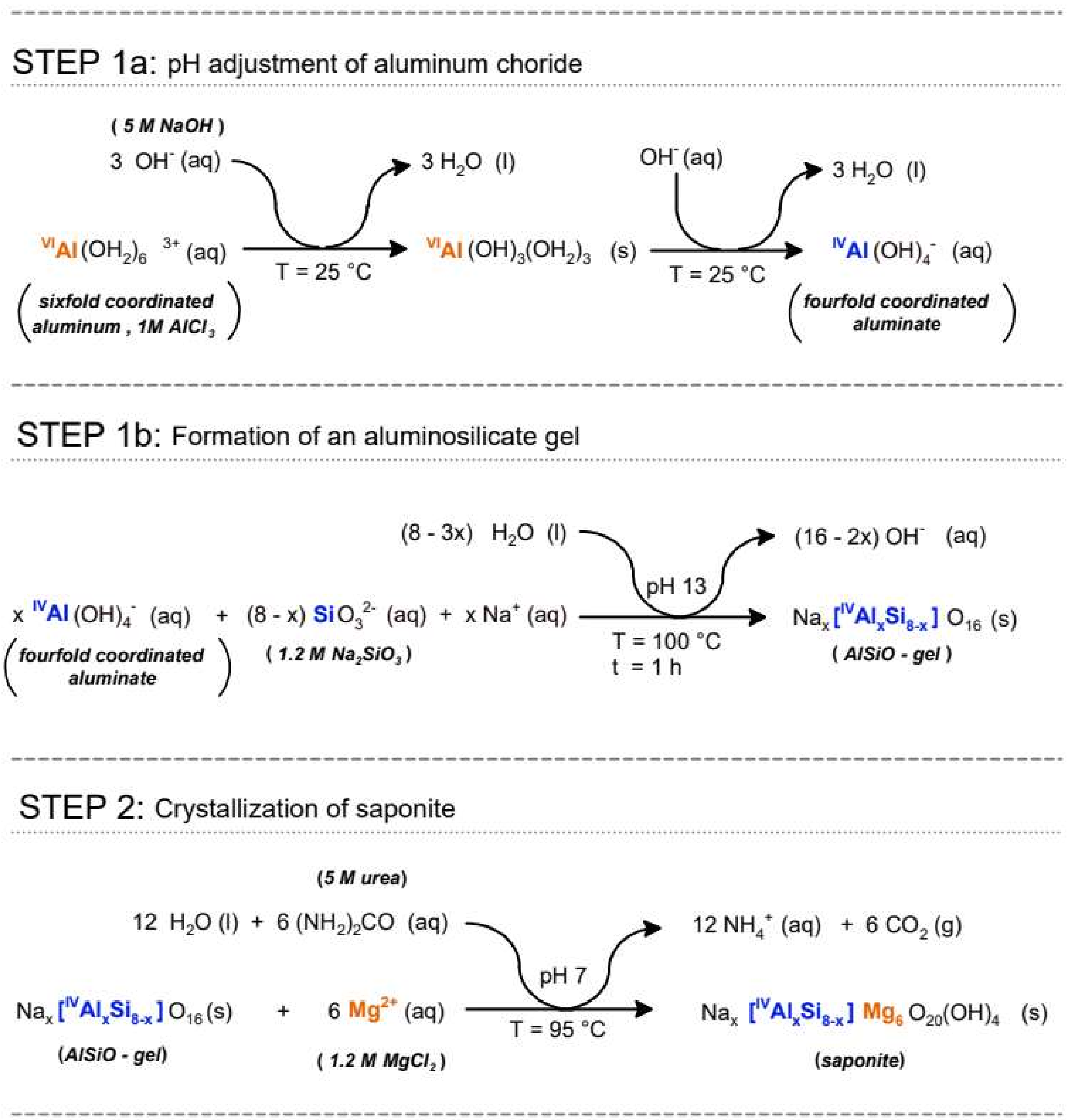

**Scheme S1.** Two-step reaction scheme for saponite synthesis: Step 1 - formation of an aluminosilicate gel: Step 1a: pH adjustment aluminum chloride and transition to a fourfold coordination, Step 1b: formation of an aluminosilicate gel from fourfold coordinated aluminate and meta-silicate ions Step 2: formation of saponite from the aluminosilicate gel and magnesium ions.

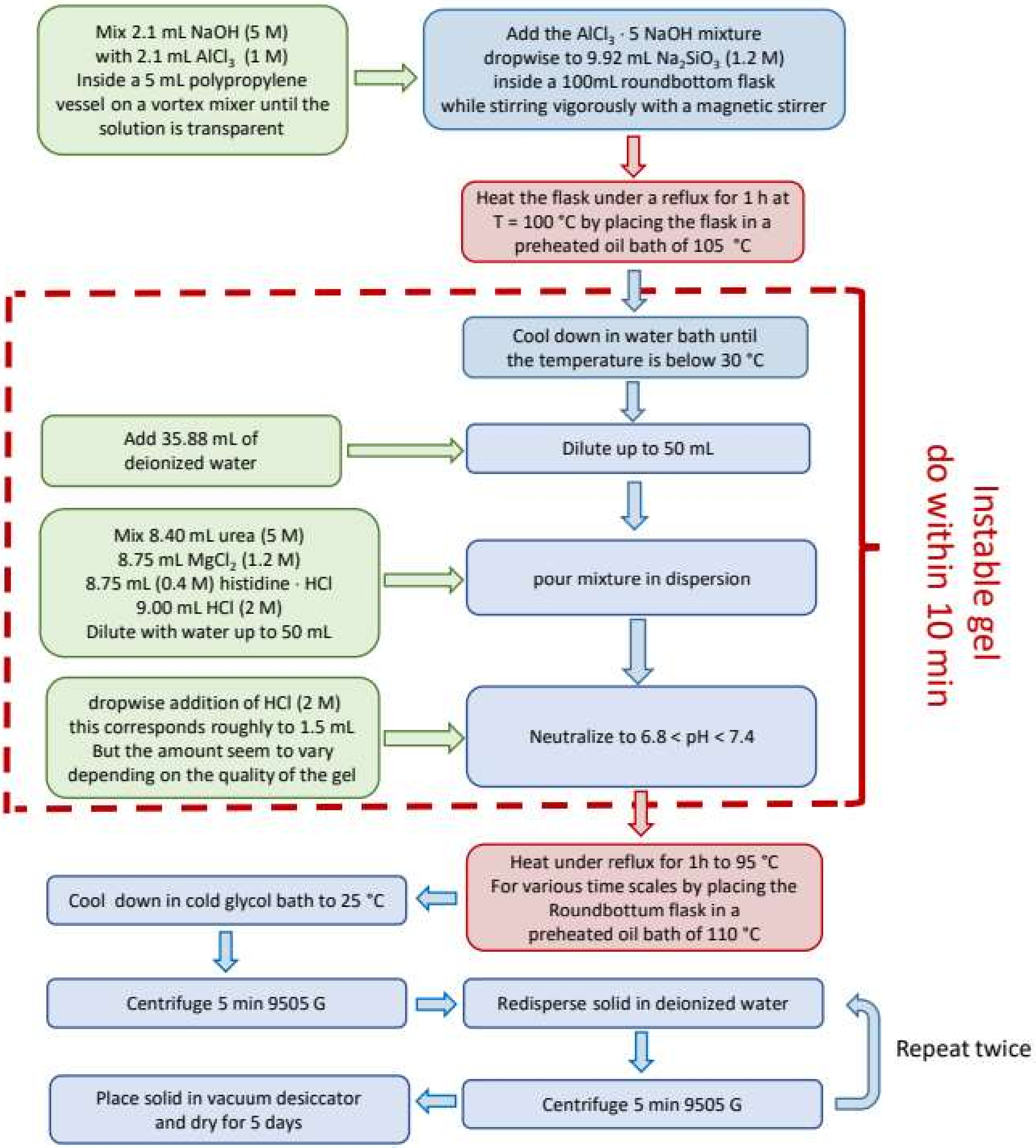

**Scheme S2.** Synthesis procedure flow diagram:

### 1.2. *Transmission electron microscopy (TEM):*

In order to image the particle size and morphology of any precipitated material during saponite synthesis, a drop of the washed suspension was placed onto a holey carbon Cu TEM grid and dried in the desiccator (10-30 mbar, 25 °C). To determine inter-layer distances within the synthesized material, an aliquot of the 24 h sample was embedded in an acryl resin (LR-White, Sigma Aldrich). For this purpose, 100 mg of the centrifuged sample material was dispersed in 1 mL of acetone and mixed in an incubator for 24 h and centrifuged afterward at 9000 rpm. This mixing step was repeated with 10/90, 30/70, 50/50, 30/70 10/90 and 100/0 resin/acetone mixtures. 6 µL of LR-white accelerator (Sigma Aldrich) was added to 1 mL of a sample/resin mixture, and in a separate vial, 12 mg benzoyl peroxide (BPO) was dissolved in 1 mL of resin. The BPO/resin mixture was added to the sample/accelerator/resin mixture, stirred vigorously, and 10 µL of this mixture was dropcast onto a TEM grid and cured for 2 days in an oven at 60 °C.

TEM images were acquired at 200 keV with an FEI Tecnai™ G2 F20 X-Twin microscope fitted with a field emission gun electron source, a Gatan Imaging Filter Tridiem™, an energy dispersive X-ray spectrometer (EDS), and a Fischione HAADF detector. From the EDS analyses the molar ratios: [Al]/[Si], [Mg]/[Si] and [Na]/[Si] ratios were evaluated and thus the four- to six-fold-coordinated aluminium molar ratios could be determined by solving mass-balance equations (Eq. S1 – S5).

### 1.3. *X-ray diffraction (XRD, HEXD):*

Diffraction analyses of the dried powders were performed with a PANanlytical Empyrean X-ray powder diffractometer, equipped with a Cu source. High-energy X-ray diffraction (HEXD) analyses of these same powders were performed at beamline 11-ID-B of the Advanced Photon Source (APS, USA) using an energy of 58.677 keV and a Mar 2D image plate detector. For the HEXD analyses, dried sample powders were filled into Kapton capillaries and data was acquired in each case for 1 minute. An empty Kapton capillary and a $CeO_2$ standard were measured for the background subtraction and calibration of the diffraction data. Data correction and azimuthal integration from 2D to 1D was performed using FIT2D.[1,2] The pair distribution function (PDF) data was generated by applying a rotation averaged Fourier transformation to the $q$-weighted data up to $q_{max}$ = 25 Å$^{-1}$ with PDFgetX3.[3] The data was analyzed with the PDFgui software[4] to optimize lattice constants anisotropic displacement parameters.

### *1.4. Fourier-Transform Infrared Spectroscopy (FTIR):*

The dried powders (1 mg) were mixed with cesium iodide (CsI, 350 mg, Sigma Aldrich) and pressed into tablets by using a uniaxial hydraulic press (98 kN, Carl Zeiss Jena). The tablets were measured in transmission mode on a Bruker Vertex 80v spectrometer, by using two beam-splitters (KBr and mylar) to cover the full range from mid (M)- to far (F)-IR (4000 – 150 $cm^{-1}$). The absorbance was measured with respect to a blank CsI tablet. The intensity of the FIR-region was normalized to the MIR region and baseline correction was done after merging both regions together. All spectra were normalized to the averaged intensity to eliminate sample weighting errors.

### *1.5. Cation exchange capacity (CEC):*

To determine the CEC, 100 mg of dried solid samples were dispersed in 100 mL DIW, the pH was adjusted to 7 (2 M HCl) and the samples were left to equilibrate for 3 d at room temperature. In the second step 0.1 to 1 mL of dispersed sample aliquots were mixed with 100 µL of 2.5 mM methylene blue (99%, Fischer Scientific) solution, diluted to 10 mL and stirred at 1000 rpm for 2 hours. Afterwards, the solids were separated from the solutions by filtration through polytetrafluoroethylene syringe filters (0.2 µm). Before use, the filters were washed three times with the sample/methylene blue mixture, to saturate the filters with methylene blue. Subsequently, the CEC was measured[5] by analyzing the samples spectroscopically over an integrated range between 650 and 675 nm (Fisher Scientific Evolution 220 Spectrometer). A series of standards was prepared for calibration using the methylene blue solutions (1, 2.5, 5, 7.5, 10, 12.5 µM).

## 2. Supporting Results and Data Analyses

### 2.1. Assignment of FTIR vibration modes during various stages of the saponite formation

The FTIR absorbance spectrum of the precursor aluminosilicate gel (spectrum I in Fig. S1) was dominated by strong antisymmetric Si-O stretching vibration modes between 950 - 1250 $cm^{-1}$ and rocking Si-O vibration modes at 445 $cm^{-1}$.[6-8] Furthermore, the band at 870 $cm^{-1}$ is characteristic for SiOH/$SO^-$ stretch vibrations,[7, 9] while the broad peak at 704 $cm^{-1}$ stems from red-shifted symmetric $[Si,Al]O_4$ vibrations, as e.g. observed for $AlO_4^-$ substituted $SiO_2$ glasses and aluminosilicate minerals (e.g., sodamelilitie, albite, and kalsilite).[9, 10] Finally, the broad peak at 580 $cm^{-1}$ likely originated from symmetric stretch vibrations of $AlO_6$ octahedra (found in e.g., jadeite[10] and mullite[11]).

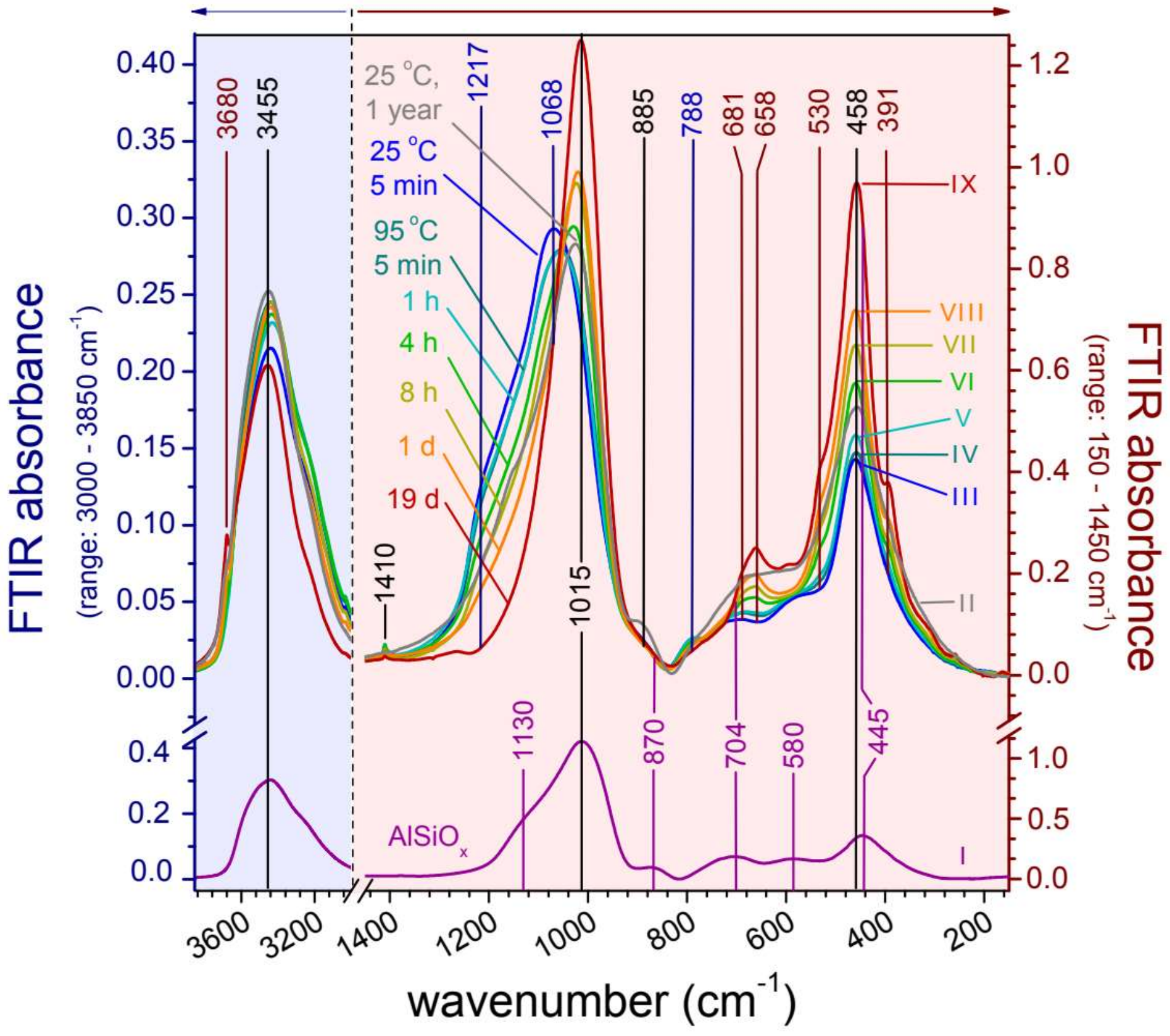

**Figure S1.** FTIR spectra of the amorphous aluminosilicate gel (bottom spectrum) and end-products, saponite, obtained after different crystallization times at 95 °C including one spectrum of a sample collected 5 min after mixing the aluminosilicate gel with the $MgCl_2$ solution at 25 °C. All assignments are listed in Table S2. Note that the two y-axes denote absorbance intensity ranges for two different parts of the spectra.

Immediately after pH neutralization and $MgCl_2$ addition but prior to heating, the FTIR pattern revealed a substantial change in the molecular bonds upon mixing of the gel with the $MgCl_2$ solution (Fig. S1, III). Compared to the aluminosilicate spectrum (Fig. S1, I), the antisymmetric Si-O and Al-O vibrations broadened and shifted to higher wavenumbers (from 1015 to 1068 $cm^{-1}$). This shift can be explained by an increase in dihedral angle of Si-O-Si bonds at a constant Si-O vibration force constant.[7, 12-14] The broadening phenomenon is characteristic for disordered silicate phases.[7, 12] and is composed of two types of vibration modes: $AS_1$ in-phase antisymmetric stretch modes (1000 - 1100 $cm^{-1}$) and $AS_2$ out-of-phase antisymmetric modes (1100 - 1200 $cm^{-1}$) Thus, overall, the addition of magnesium seems to increase the dihedral angle of Si-O-Si bonds and/or leads to a more disordered structure enabling $AS_2$ vibrations.

**Table S2.** FTIR assignments for precursor aluminosilicate gel, the starting amorphous magnesium precursor and the final saponite after 19 d of reaction at 95 °C.

| **Peak position ($cm^{-1}$)** | **$AlSiO_x$ gel** | **MgAlSiOx amorphous 5 min at 25 °C** | **Saponite after 19 d @ 95 °C** | **Assignment(s)** | |
|---|---|---|---|---|---|
| 391 | | | X | $\nu$ $[AlO_6]$ [15] | |
| 445 | X | X | X | $b_1^4$ $[Si_2O_7]$ [16]<br>$b_2^5$ $[Si_2O_7]$, [16]<br>$\nu_4 (a_1)$ $[SiO_4]$ [17] | Si-O bending modes [6, 16] |
| 458 | | X | X | $\delta_T$ [(Al)OH] [6, 17, 18] | AlOH translation [6, 17, 18] |
| 530 | | | X | $\nu_6 (f_{2u})$ [Mg-OH] [17, 18] | perpendicular deformation [17, 19] |
| 580 | X | | | $\nu_s$ $[Al^{VI}O_6]$ [19, 20] | Condensed $AlO_6$ octahedral [22] |
| 658 | | | X | $\delta_L$ [(Mg)OH] [16, 23] | MgOH liberation [16] |
| 681 | | | X | $\nu_2 (a_1)$ $[SiO_4]$ [17, 24]<br>$a_1^2$ $[Si_2O_7]$ [16, 21] | symmetric Si-O vibrations [17, 24] |
| 704 | X | X | | $\nu_s$ $[SiO_4 \cdot AlO_4]$ [10, 25] | Associated with network $\nu AlO_4$ substituted silicate glasses [10, 25] |
| 788 | | X | | $\nu_s$ $[SiO_4]$ [6, 17] | |
| 870 | X | | | $\nu_s$ $[SiOH/SO^-]$ [7, 9] | $SiOH/SO^-$ stretch vibrations [7, 9] |
| 915 | | X | X | $\delta$ $[(Al^{VI})_2OH]$ [`16, 23, 26] | OH bending from octahedral $Al^{VI}$ [23] |
| 1015, 1068 | X | X | X | $\nu_3 (e_1)$ $[SiO_4]$ [17, 24]<br>$\nu_1 (a_1)$ $[SiO_4]$ [17, 24]<br>$b_1^1$ $[Si_2O_7]$ ,$b_2^2$ $[Si_2O_7]$ [16] | in-phase antisymmetric Si-O and Al-O stretching modes ($AS_1$) [7, |
| 1130 | X | | | $\delta$ (Si)OH [27] | |
| 1217 | | | X | $\nu_{as}$ $SiO_4$ [6, 7, 12] | out-of-phase antisymmetric SiO stretching modes ($AS_2$) [7, 12] |
| 1410 | | X | X | $\delta$ $[NH_4^+]$ [28] | $NH_4$ bending [28] |
| 3000-3650 | X | X | X | $\nu$ $[OH (H_2O)]$ [23, 28]<br>$\nu$ $[NH_4]$ [28] | neutral hydroxyl stretch vibrations [23, 28] |
| 3590 | | | ? | $\nu$ [(Si,Al)OH] [29] | acidic OH stretch vibrations [29] |
| 3680 | | | X | $\nu$ $[(Mg)_3OH]$ [20, 23, 29] | |

X = vibration is observed in spectrum.

? = This vibration is likely present, but it is difficult to verify due to strong overlap with the neighboring vibration.

During the transformation of the amorphous material towards a trioctahedral, 2:1 layered clay structure, the spectra contained characteristic Mg-OH liberation modes (658 $cm^{-1}$) and the acidic Mg-OH stretch vibrations (3680 $cm^{-1}$, Fig. S1). The shoulder at 1217 $cm^{-1}$ of antisymmetric Si-O vibration modes diminished and the remaining peak narrowed due to a reduced variation in the dihedral Si-O-Si bond angle, and reduced contribution of $AS_2$ modes due to an increased degree of structural order.[7, 12-14] The degree of $^{IV}Al$ substitution likely also had some effect on the position of the dominant $\nu_{as}$(Si-O) mode, however, the effect had been shown to be small in saponite compared to talc (both around 1018 $cm^{-1}$).[16] On the other hand, the previously reported comparison among several phylosilicates[16] suggested that geometrical and/or charge effects had a more pronounced effect on the position of the $\nu_{as}$ (Si-O) mode as compared to $^{IV}Al$- substitution.
With increasing time several of the vibrations narrowed due to increased structural order (i.e. $\nu_{as}$(Si-O), δ(SiO) and (Mg)OH liberation) and the FTIR spectrum of the material synthesized after 19 d contained several vibration modes that were characteristic for a typical trioctahedral smectite (Table S2): the MgOH perpendicular deformation (530 $cm^{-1}$), the (Mg)OH liberation (658 $cm^{-1}$) and the (Mg)OH stretch vibration (3680 $cm^{-1}$). The (Mg)OH liberation was clearly red-shifted with respect to talc, which is characteristic for Mg-rich trioctahedral smectite clays: saponite and hectorite.[16] Apical $AlO_4$ vibrations at ≈ 840 $cm^{-1}$ were not observed, however, this vibration is known to be rather weak and was only observed for highly-crystalline saponite synthesized at temperatures above 200 °C.[19] The sample at 25 °C/1 y shares many spectral features with the one for 4 h at 95 °C, except for the increased absorbances at 580, 704 and 885 $cm^{-1}$ in the former one. As mentioned above, the 580 $cm^{-1}$ is associated with symmetrical $AlO_6$ vibrations that may originate from a gibbsite impurity as suggested by the XRD measurements.

### 2.2. Compositional analysis, mass balances and cationic exchange capacity:

Saponite is a smectite type phyllosilicate mineral with brucite-like $Mg(OH)_2$-layer in between two layers of six-membered silica tetrahedral, where two-thirds of the OH- ions (in the brucite layer) are replaced by $O^{2-}$ ions. These $O^{2-}$ ions are interconnected with neighboring sheets of tetrahedral aluminosilica (Fig. S1). The combined structure is a trioctahedral phyllosilicate-like talc, where practically all octahedral positions are occupied (≫2/3). Some $^{IV}Si^{4+}$ are replaced by $^{IV}Al^{3+}$ ions, which leaves an excess negative charge that is counterbalanced by the metal cations in between the tetrahedral-octahedral-tetrahedral (TOT) sandwich sheets (Fig. 3b, main text). However, some aluminum may occupy octahedral lattice positions as described in more detail below.

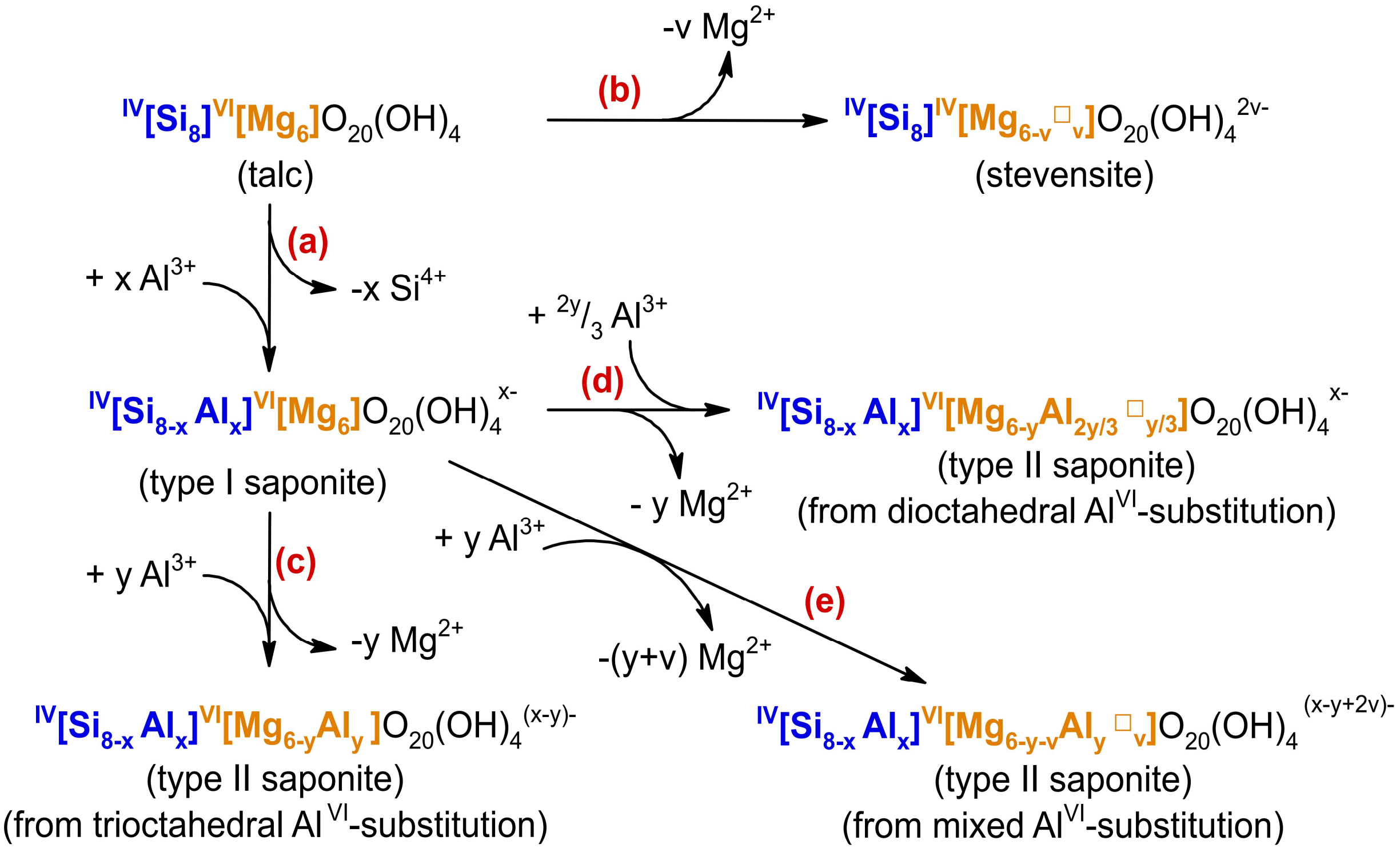

**Scheme S3. Illustration how various $Al^{3+}$ substitution mechanisms affect the charge balance of the trioctahedral clay structures, where: (a) talc to type-I saponite substitution; (b) talc to stevensite substitution, (c-e) several type-I to type-II saponite substitution schemes: (c) trioctahedral, (d) dioctahedral and (e) mixed octahedral $^{VI}Al$ substitution).**

Saponite can be subdivided (Scheme S3 a) to: 'ideal/type I' saponite where the $Al^{3+}$ solely replaces 4-fold-coordinated $^{IV}Si^{4+}$ ions; and 'non-ideal/type II' saponite where $Al^{3+}$ replaces both 6-fold-coordinated $^{VI}Mg^{2+}$ and 4-fold-coordinated $^{IV}Si^{4+}$. The 4-fold/6-fold Al ratio ($^{IV}Al/^{VI}Al$) is known to increase for increasing reaction temperatures[30] and decreasing water contents in the starting aluminosilicate gel.[31] Moreover, isomorphous substitution of $^{VI}Mg^{2+}$ by $^{VI}Al^{3+}$ may occur through two different mechanisms: trioctahedral or equimolar $^{VI}Al$ -substitution where every $^{VI}Mg^{2+}$ is replaced by one $^{VI}Al^{3+}$ (Scheme S3 c); or a dioctahedral $^{VI}Al^{3+}$ -substitution where 3 $Mg^{2+}$ ions are replaced by 2 $Al^{3+}$ and 1 vacancy (□) so that the structure shifts towards dioctahedral beidelite with cation vacancies (Scheme S3 d). Such vacancies can also exist in the absence of aliminium as it is the case for stevensite (Scheme S3 b). The catalytically active Bronstead or Lewis acid sites are formed in the presence of 4-fold-$O^{2-}$-coordinated $^{IV}Al^{3+}$ ions that are substituted within the silicate layer. This makes them suitable for various synthesis processes including dehydration, isomerization, substitution and oxidation reactions.[32-34] Moreover, the surface area and acidity of these layered silicates can be tuned by substitution of transition metal cations into the structure.[32, 35-36]

In the case of trioctahedral $^{VI}Al$-substitution, all octahedral cation positions remain occupied and the octahedral layer becomes positively charged (reducing the net negative charge of the TOT-layered sheet), whereas in the case of dioctahedral $^{IV}Al$ substitution the octahedral layer remains charge-neutral. Cation exchange measurements revealed a decreasing cation exchange capacity for increasing $^{VI}Al$ which supports the trioctahedral $^{VI}Al$ substitution mechanism.[37] On the other hand saponite phases are known to exist with $[^{IV}Al] > [^{VI}Al]$ where exchangeable interlayer cations are only possible if the structure contains vacancies, since the TOT-sheet would otherwise be positively charged. Both substitution mechanisms may occur in parallel leading to a mixed substitution mechanism (Scheme S3 e). The extent to which the $Al^{3+}$ occupies either octahedral or tetrahedral positions plays a key role for the crystallinity, cation exchange, capacity and presence of catalytically active acidic sites. Ideally, $Al^{3+}$ would only occupy tetrahedral sites, which are catalytically active and increase the cation exchange capacity, whereas octahedral $Al^{3+}$ ions are catalytically inactive and have a reduced the cation exchange capacity that leads to increased structural disorder.[29, 30-31, 36,]

For our experiments, the fraction of $^{IV}Al^{3+}$ and $^{VI}Al^{3+}$ was derived experimentally. Since, the structure of saponite is known (Fig. 3b, main text), and the atomic compositions were derived by TEM-EDS, the fraction of both tetrahedral $^{IV}Al^{3+}$ and octahedral $^{VI}Al^{3+}$ can be resolved by solving the charge balance, lattice position balance and mass balance equations. By assuming saponite has a composition as illustrated in Fig. 3b (main text), the following charge balance is derived:

$$\boldsymbol{z = 2v + x - y}$$

**(Eq. S1)**

Where $z$, $v$, $x$ and $y$ represent molar ratios of sodium ions, octahedral vacancies, tetrahedral $^{IV}Al^{3+}$ and octahedral $^{VI}Al^{3+}$ ions as illustrated by the saponite structure and substitution scheme (Fig. 3b (main text), Scheme S2). From its structure we can derive the following lattice position mass balances:

$$\frac{[Al]}{[Si]} = \frac{x+y}{8-x} \quad \textbf{(Eq. S2)} \qquad \frac{[Na]}{[Si]} = \frac{z}{8-x} \quad \textbf{(Eq. S3)} \qquad \frac{[Mg]}{[Si]} = \frac{6-y-v}{8-x} \quad \textbf{(Eq. S4)}$$

The molar ratios: [Al]/[Si], [Mg]/[Si] and [Na]/[Si] were measured by TEM-EDS analyses of multiple samples and from this data the molar ratio for the tetrahedral aluminum positions, $x$, was estimated from:

$$x = \frac{8r-12}{r+4}, \text{ where } \quad r = 3\frac{[Al]}{[Si]} + 2\frac{[Mg]}{[Si]} + \frac{[Na]}{[Si]} \qquad \textbf{(Eq. S5)}$$

The results of the compositional analysis and cationic exchange capacity, were summarized in two tables Table S3 for the amorphous precursors and Table S4 for nanocrystalline saponite. The lattice position balances (S2 – S4) are only valid for samples with a (nanocrystalline) saponite structure (Table S4) and consequently we could only calculate the atomic ratios v, x, y, z for those samples.

**Table S3.** Composition of the amorphous precursor aluminosilicate gel (end product of Stage 1) with molar ratios from TEM-EDS per $[Si_{8-x}Al_x]$ unit and 5 min after mixing with the $MgCl_2$ solution and initiation of heating.

| | **AlSiOx (95 °C after 1 h)** | | | **AlSiMgOx** |
|---|---|---|---|---|
| | core | **rim** | **total area** | **5 min after reaching 95 °C** |
| Na | 0 | 0.88 (5) | 0.35 (4) | 0.17 (7) |
| Si | 7.39 (17) | 7.04 (15) | 7.15 (18) | 7.58 (18) |
| Al | 0.61 (5) | 0.95 (6) | 0.85 (7) | 0.42 (19) |
| Mg | | | | 0.81 (17) |
| O | 27.9 (4) | 26.8 (3) | 27.4 (5) | 21.6 (7) |

1) Number in between brackets represents an error margin of the last digit(s), which was based on standard deviation of three EDX measurements.
2) All atomic ratios were normalized to $r_{Si} + r_{Al} = 8$.

Table S4. Saponite composition for samples produced at 95 °C after 24 h, and at 25 °C after 19 d with molar ratios from TEM-EDS and CEC per $O_{20}(OH)_4$ unit.

| Atom type | Variable | 24 h | 19 d |
|---|---|---|---|
| Mg | $r_{Mg}$ = 6-*y*-*v* | 5.04 (17) | 5.3 (4) |
| Si | $r_{Si}$ = 8-*x* | 7.55 (15) | 7.3 (4) |
| Na | $r_{Na}$ = *z* | 0.16 (8) | 0.33 (4) |
| Al | $r_{Al}$ = *x*+*y* | 1.18 (13) | 1.25 (19) |
| $^{IV}$Al | $r^{IV}_{Al}$ = *x* | 0.45 (5) | 0.68 (17) |
| $^{VI}$Al | $r^{VI}_{Al}$ = *y* | 0.74 (8) | 0.56 (11) |
| □ | $r_{\square}$ = *v* | 0.23 (11) | 0.12 (9) |
| $H_2O$ | $r_{H2O}$ = *n* | 2.1 (7) | 2.8 (8) |
| **variable** | **unit** | | |
| Mw | (g/mol) | 703 (9) | 716 (10) |
| CEC | (mmol/g) | 0.15 (2) | 0.42 (2) |
| CEC | (mol/mol) | 0.11 (2) | 1.30 (2) |

1) Number in brackets represents an error margin of the last digit(s), which was based on standard deviation of three EDX measurements.
2) All atomic ratios were normalized to obey Eqs. S1-S5 above, such that: $r_{Si} + r_{Al} + r_{Mg} + r_{\square} = (8-x) + (x+y) + (6-y-v) + v = 14$.
3) The molar $H_2O$ ratio was estimated from the O(K) EDX signal intensity

### 2.3. High Energy X-ray Diffraction patterns:

The high energy X-ray diffraction patterns after beamline corrections, calibration, azimuthal integration and subtraction of Kapton capillary background, which were quenched during various stages of the reaction were shown in Figure S2.

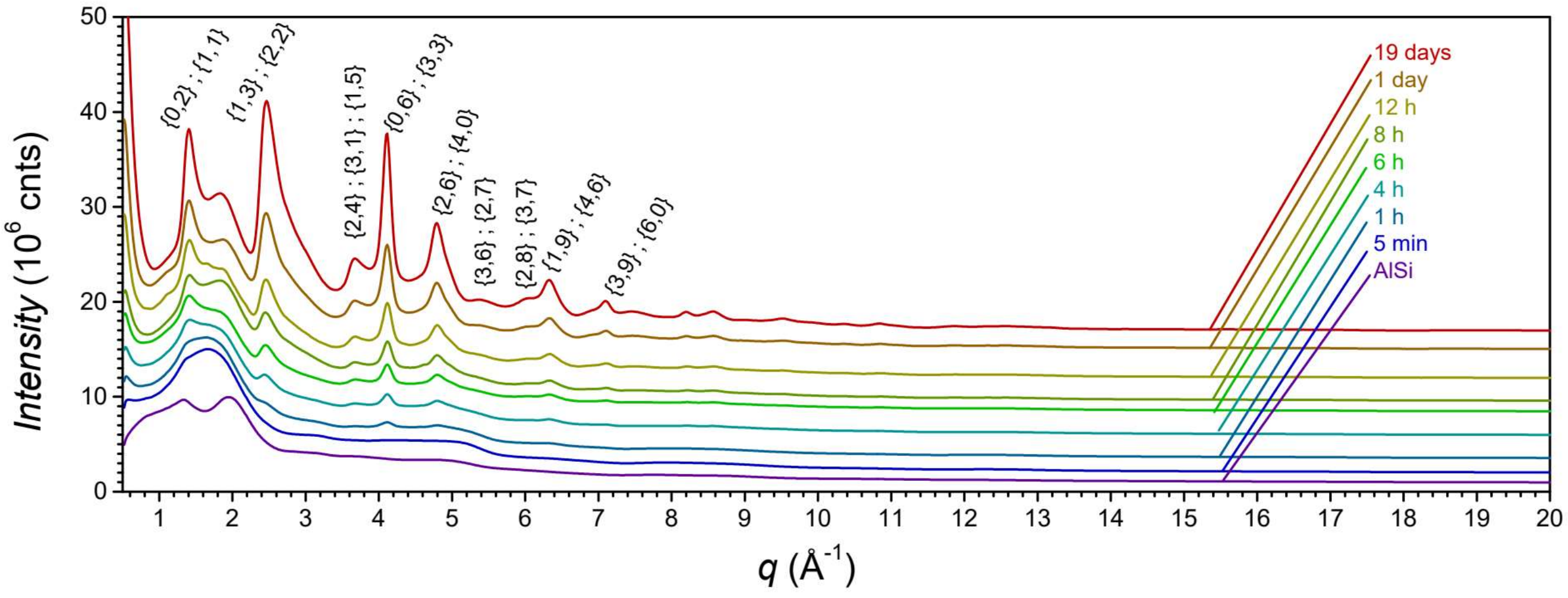

**Figure S2**. High energy X-ray diffraction patterns at various stages of the saponite formation process at 95 °C.

### 3.4. **Kinetics of saponite formation:**

In order to study the effect of magnesium concentration on saponite formation we tested three possible scenarios to evaluate the dependency of the magnesium concentration on kinetic rates. Saponite was synthesized with two [Mg]/[Si] ratios: 60/68 (stoichiometric ratio) and 120/68 (double the stoichiometric ratio. In all three scenarios, an aluminosilicate gel was formed prior to the formation of saponite. By assuming that the aluminosilicate gel has the same Na:Al:Si ratio as an ideal type-I saponite, its idealized formation reaction can be written as:

$$\mathbf{0.2\,Na^+ + 0.2\,Al(OH)_4{}^- + 1.13\,SiO_3{}^{2-} + 0.73\,H_2O \rightarrow Na_{0.2}Al_{0.2}Si_{1.13}O_{2.67} + 2.26\,OH^-}$$

**(Eq. S6)**

***Scenario 1:***

Let us assume that aluminosilicate gel reacts with magnesium to saponite without intermediate phases according to the following equation (Eq. S7):

$$Na_{0.2}Al_{0.2}Si_{1.13}O_{2.67} + Mg^{2+} + (NH_2)_2CO + 2.33H_2O \rightarrow saponite + CO_2 + 2\ NH_4^+$$

**(Eq. S7)**

Here, the atomic ratio in saponite is normalized against magnesium: $Na_{0.2}Al_{0.2}Si_{1.13}Mg_1O_{3.33}(OH)_{0.67}$ and all aluminum ions are assumed to occupy the tetrahedral positions. For the sake of simplicity we ignore the incorporation of 6-fold-coordinated $^{VI}Al^{3+}$ and ion vacancies. Since the aluminum concentration is small in comparison with the silicon and magnesium concentrations, and because saponite can also be formed with smaller aluminum fractions, we do not expect that the $^{VI}Al^{3+}/^{IV}Al^{3+}$ may have a strong effect on the reaction order. For an equimolar aluminosilicate-to-magnesium ratio, the rate of aluminosilicate depletion should be equal to the rate of magnesium ion depletion. Moreover, the high excess of urea concentration will not affect the reaction rate. Consequently, the rate of saponite formation, which also correlates with the rate of magnesium depletion, can be described as a second order reaction:

$$\frac{d[saponite]}{dt} = -\frac{d[Mg^{2+}]}{dt} = k[Mg^{2+}][AlSiO_x] = k[Mg^{2+}]^2$$

**(Eq. S8)**

However, several magnesium ions can react with a single aluminium silicate cluster, whose average size is probably larger than [$Na_{0.2}Al_{0.2}Si_{1.13}O_{2.67}$]. Consequently, the reaction is more likely described by the following differential equation:

$$\frac{d[saponite]}{dt} = -\frac{d[Mg^{2+}]}{dt} = k_1[Mg^{2+}]^m[AlSiO_x] = k[Mg^{2+}]^n$$

**(Eq. S9)**

Here, *m* is the average number of magnesium ions that reacts with a single aluminosilicate cluster and *n* is the order of the reaction ($n = m+1$). Consequently, the progression of the saponite formation process is described by:

$$\alpha = \frac{[saponite](t)}{[saponite]_{t=\infty}} = \frac{[Mg^{2+}]_{t=0}}{[saponite]_{t=\infty}} \cdot \left( 1 - \sqrt[n-1]{\frac{1}{\left(1 - \frac{([Mg^{2+}]_{t=0})^{n-1} \cdot k \cdot t}{n-1}\right)}} \right)$$

**(Eq. S10)**

In the above case, when magnesium ions reacts with aluminosilicate to saponite, either a second reaction order (Eq. S8) or higher ($n^{th}$) reaction order with $n > 2$ could be expected (Eq. S10).

***Scenario 2:***

In the 2nd scenario, we assume that the condensation of magnesium to a brucite ($Mg(OH)_2$)-like layer before saponite could be formed, is the rate-limiting stage according to the following reaction equations:

$$Mg^{2+} + (NH_2)_2CO + 3H_2O \rightarrow Mg(OH)_{2\ (brucite\ layer)} + CO_2 + 2\ NH_4^+$$

**(Eq. S11)**

In this case, the brucite reacts with the aluminosilicate gel to crystalline saponite:

$$Na_{0.2}Al_{0.2}Si_{1.13}O_{2.67} + Mg(OH)_{2\ (brucite\ layer)} \rightarrow saponite + 0.67H_2O$$

**(Eq. S12)**

If we assume that brucite layer formation is the rate-limiting stage, then the rate of saponite formation becomes 1st order according to the following differential equation:

$$\frac{d[saponite]}{dt} = \frac{d[Mg(OH)_2]}{dt} = -\frac{d[Mg^{2+}]}{dt} = k[Mg^{2+}]$$ **(Eq. S13)**

This results in the following rate of saponite formation:

$$\alpha = \frac{[saponite](t)}{[saponite]_{t=\infty}} = \frac{[Mg^{2+}]_{t=0}}{[saponite]_{t=\infty}} \cdot \left(1 - \exp(-kt)\right) \quad \textbf{(Eq. S14)}$$

***Scenario 3:***

In the 3rd scenario, we assume that magnesium immediately condenses with the aluminosilicate gel to form an magnesium aluminosilicate that remains amorphous, but it has the same stoichiometry as saponite according to:

$$Na_{0.2}Al_{0.2}Si_{1.13}O_{2.67} + Mg^{2+} + (NH_2)_2CO + 2.33H_2O \rightarrow$$
$$Na_{0.2}Al_{0.2}Si_{1.13}MgO_{3.33}(OH)_{0.67}\ (amorphous) + CO_2 + 2\ NH_4^+$$

**(Eq. S15)**

This initial fast condensation is followed by a rate-limiting crystallization process:

$$Na_{0.2}Al_{0.2}Si_{1.13}MgO_{3.33}(OH)_{0.67}\ \ (amorphous) \rightarrow saponite \quad \textbf{(Eq. S16)}$$

In this latter case, saponite formation also follows a 1st order process but it depends on the amorphous magnesium aluminosilicate concentration. Thus, the rate of saponite formation, $\alpha$, can be described by:

$$\alpha = \frac{[saponite](t)}{[saponite]_{t=\infty}} = \frac{[MgAlSiO_x]_{t=0}}{[saponite]_{t=\infty}} \cdot \left(1 - exp(-kt)\right)$$

**(Eq. S17)**

In this case, an excessive magnesium concentration with respect to the constant aluminosilicate concentration will not affect the initial amorphous magnesium aluminosilicate concentration $[MgAlSiO_x]_{t=0}$. It is reasonable to assume that the initial gelation (Eq. S15) is much faster than the final crystallization (Eq. S16). Consequently, the gelation will barely affect the rate of saponite formation. On the other hand, in the case of the brucite-layer limited reaction (scenario 2, Eq. S14), this stage directly depends on the magnesium concentration. In such a case, excess magnesium would increase the rate of saponite formation.

The obtained reaction rate constants for a first order reaction of saponite formation as several temperatures, are shown in the Table S5:

**Table S5.** Table of reaction rate constants.

| Formation rate based on | | *k* ($10^{-6}$ $s^{-1}$) |
|---|---|---|
| XRD peak intensities | $Mg_6$ sap at 25 °C | 0.007 (3) |
| | $Mg_6$ sap at 60 °C | 0.19 (7) |
| | $Mg_6$ sap at 95 °C | 21 (4) |
| | $Mg_{12}$ sap at 95 °C | 23 (2) |
| Intensities of PDF correlation distances | Na-O at d = 2.38 Å | 11 (1) |
| | Oc-O at d = 2.06 Å | 7.0 (4) |
| | T-(Oc)-T at d = 8.18 Å | 18 (1) |
| Shift of PDF peak position | T-O | 25 (5) |
| Error margins of the last digit are listed in between brackets | | |